\documentclass[11pt,a4paper]{article}

\usepackage{jcappub}
\usepackage{graphicx}
\usepackage{amssymb}
\usepackage{amsmath}
\usepackage{bm}
\usepackage{latexsym}
\usepackage{epsfig}
\usepackage{psfrag}
\usepackage{ulem}
\usepackage{textcomp}
\usepackage{color}
\usepackage{pstricks}
\usepackage[utf8]{inputenc}

\makeatletter
\gdef\@fpheader{}
\makeatother

\def\nn{\nonumber} 
\def\pa{{\partial}}
\def\f{\frac}
\def\l{\left}
\def\r{\right}
\def\d{{\rm d}}
\def\Mpl{M_{_{\rm Pl}}}
\def\beq{\begin{equation}}
\def\eeq{\end{equation}} 
\def\beqa{\begin{eqnarray}}
\def\eeqa{\end{eqnarray}} 

\def\bA{\bar A}
\def\cA{\mathcal A}

\def\psb{{\mathcal P}_{_{\rm B}}}

\def\vk{{\bm k}}

\def\vkb{{\bm k}_{2}}
\def\vkc{{\bm k}_{3}}
\def\ka{k_{1}}
\def\kb{k_{2}}
\def\kc{k_{3}}

\def\Mp{M_{_{\rm Pl}}}
\def\cG{{\cal G}}
\def\cP{{\cal P}}

\def\ee{\eta_{\rm e}}

\def\a1{\alpha_{_1}}
\def\b1{\beta_{_1}}
\def\de1{\delta_{_1}}
\def\g1{\gamma_{_1}}
\def\bnl{b_{_{\rm NL}}}
\def\fnl{f_{_{\rm NL}}}

\def\cH{{\mathcal H}}

\newcommand{\viz}{\textit{viz.~}}
\newcommand{\ie}{\textit{i.e.~}}

\begin{document}
\title{Enhancing the cross-correlations between magnetic 
fields and scalar perturbations through parity violation}
\author{Debika Chowdhury$^\dag$,}
\emailAdd{debika@physics.iitm.ac.in}
\affiliation{$^\dag$Department of Physics, Indian Institute of 
Technology Madras, Chennai~600036, India}
\author{L.~Sriramkumar$^\dag$,}
\emailAdd{sriram@physics.iitm.ac.in}
\author{Marc Kamionkowski$^\ddag$} 
\affiliation{$^\ddag$Department of Physics and Astronomy, Johns Hopkins 
University, 3400 N. Charles Street, Baltimore, MD~21218, U.S.A.}
\emailAdd{kamion@jhu.edu}


\abstract
{One often resorts to a non-minimal coupling of the electromagnetic field 
in order to generate magnetic fields during inflation.
The coupling is expected to depend on a scalar field, possibly the 
same as the one driving inflation.
At the level of three-point functions,
such a coupling leads to a non-trivial cross-correlation between the 
perturbation in the scalar field and the magnetic field.
This cross-correlation has been evaluated analytically earlier for the 
case of non-helical electromagnetic fields.
In this work, we {\it numerically}\/ compute the cross-correlation for 
helical magnetic fields.
Non-Gaussianities are often generated as modes leave the Hubble radius.
The helical electromagnetic modes evolve strongly (when compared to the 
non-helical case) around Hubble exit and one type of polarization is 
strongly amplified immediately after Hubble exit.
We find that helicity considerably boosts the amplitude of the dimensionless
non-Gaussianity parameter that characterizes the amplitude and shape of the 
cross-correlation between the perturbations in the scalar field and the 
magnetic field.
We discuss the implications of the enhancement in the non-Gaussianity parameter
due to parity violation.}
\keywords{Primordial magnetic fields, inflation, non-Gaussianities}
\maketitle


\section{Introduction}

The observations of widely prevalent magnetic fields in the universe call 
for investigations into their origin. 
The amplitude of the magnetic fields vary over an extensive range, from a 
few micro Gauss in stars and galaxies, to around $10^{-17}$ Gauss in the 
intergalactic medium~\cite{Neronov:1900zz,Tavecchio:2010mk,Dermer:2010mm,
Vovk:2011aa,Tavecchio:2010ja,Dolag:2010ni,Taylor:2011bn,Takahashi:2011ac,
Huan:2011kp,Finke:2015ona} and the large scale 
structure~\cite{Grasso:2000wj,Widrow:2002ud,Kandus:2010nw,Widrow:2011hs,
Durrer:2013pga,Subramanian:2015lua}.
Additionally, observations of the anisotropies in the
Cosmic Microwave Background (CMB) by Planck and POLARBEAR have led to an 
upper bound of a few nano Gauss on the magnetic fields at scales of 
$1\,{\rm Mpc}$~\cite{Ade:2015cva,Ade:2015cao}.
It has been realized that such large scale magnetic 
fields may need to be generated primordially, which can then be 
amplified by astrophysical processes.
For instance, magnetohydrodynamic (MHD) processes such as the dynamo 
mechanism in astrophysical systems necessarily require a seed field 
that can act as the progenitor of the observed magnetic 
fields~\cite{Kandus:2010nw,Widrow:2011hs,
Durrer:2013pga,Subramanian:2009fu,Giovannini:2017rbc,Subramanian:2015lua}.
It has been conjectured that primordial magnetogenesis, \ie generation of 
magnetic fields via quantum fluctuations in the early universe, can produce 
the precursor seed fields which can, over the course of time, source the 
large scale magnetic fields.

\par

Inflation, on account of being the most appealing paradigm to describe the 
early universe, has also garnered tremendous interest as the framework for 
primordial magnetogenesis.
However, the rapid decay of magnetic fields generated by the standard
electromagnetic action in an expanding universe compels one to look 
for a means to circumvent this issue.
One of the feasible ways to engender magnetic fields of appropriate 
strengths seems to be the introduction of a non-minimal coupling term 
in the action~\cite{Turner:1987bw,Ratra:1991bn}.
Magnetic fields generated through breaking the conformal invariance of 
the standard electromagnetic action, due to the presence of a non-minimal 
coupling term, and with a suitable choice of the parameters involved,
have been shown to be of the pertinent amplitude and correlation length 
to be in concordance with the observations~\cite{Bamba:2003av,Bamba:2006ga,
Demozzi:2009fu,Martin:2007ue,Campanelli:2008kh,Kanno:2009ei,
Subramanian:2009fu,Urban:2011bu,Durrer:2010mq,Byrnes:2011aa,Jain:2012jy, 
Kahniashvili:2012vt,Cheng:2014kga,Bamba:2014vda,Fujita:2015iga,
Campanelli:2015jfa,Fujita:2016qab,Tsagas:2016fax,Sharma:2017eps}.

\par

In the context of primordial magnetogenesis, another interesting aspect is 
to study the magnetic fields generated due to the addition 
of a parity violating term to the standard electromagnetic action.
Such a term would lead to the generation of the so-called helical magnetic 
fields~\cite{Campanelli:2008kh,Durrer:2010mq,Jain:2012jy,Caprini:2014mja,
Sharma:2018kgs}.
In this situation, two modes with positive and negative helicity are generated,
which evolve differently and, as a consequence, can conceivably lead to 
distinct imprints, such as correlations between B-mode and E-mode polarizations 
or the temperature and B-mode polarizations in the CMB~\cite{Caprini:2003vc,
Ballardini:2014jta,Durrer:2010mq}.
They can also lead to the production of helical gravitational waves with 
possible observational imprints 
(in this context, see, for example, Refs.~\cite{Caprini:2003vc,
Ballardini:2014jta,Seto:2008sr,Durrer:2010mq}).
Further, it has been shown that, the helical fields evolve strongly in the cosmic 
MHD plasma through an inverse cascade mechanism, resulting in an augmentation of 
the power on large scales~\cite{Brandenburg:2004jv,Banerjee:2004df,Campanelli:2007tc}.

\par

An additional way to constrain these magnetic fields would be to study their
cross-correlations at the level of the three-point functions with the scalar 
perturbations and their possible observational imprints.
While such three-point functions have already been studied for the case of 
non-helical magnetic fields~\cite{Caldwell:2011ra,Motta:2012rn,Jain:2012ga,
Jain:2012vm}, we believe it is of utmost interest to study the non-Gaussianities 
produced due to the helical fields as well.
Analytically, the evaluation of these three-point functions seems to be a 
formidable task, due to the non-trivial form of the helical modes involving 
the Coulomb functions~\cite{Durrer:2010mq,Jain:2012jy,Caprini:2014mja}.
In this work, we shall {\it numerically}\/ evaluate the three-point function
involving the helical magnetic fields and the perturbations in an auxiliary 
scalar field and examine its implications.

\par

This paper is structured as follows.
In the next section, we shall discuss the action governing non-minimally
coupled helical electromagnetic fields, their quantization and the power 
spectrum associated with the magnetic field.
In Sec.~\ref{sec:psbi}, working with a specific form of the non-minimal
coupling, we shall revisit the evaluation of the power spectrum of 
helical magnetic fields arising in de Sitter inflation.
In Sec.~\ref{sec:cc}, by suitably perturbing the action, we shall arrive 
at the Hamiltonian describing the interaction between the perturbed scalar
field and the electromagnetic field.
Using the interaction Hamiltonian, we shall arrive at the formal structure 
of the three-point function describing the cross-correlation between the
perturbation in the scalar field and the electromagnetic field.
In Sec.~\ref{sec:ne}, we shall outline the numerical procedure that we 
shall adopt to compute the cross-correlation.
In Sec.~\ref{sec:bnl}, we shall first compare our numerical results with 
the analytical results available in the non-helical case for two different
situations (one leading to a power spectrum with a tilt and another which
leads to a scale invariant spectrum) and present the corresponding results 
in the helical case.
We shall conclude in Sec.~\ref{sec:c} with a brief summary of the results
we have obtained.

\par

Note that, we shall work with natural units such that $\hbar=c=1$ and set 
the Planck mass to be $\Mpl=\l(8\,\pi\, G\r)^{-1/2}$. 
We shall adopt the metric signature of $\l(-, +, +, +\r)$. 
Greek indices shall denote the spacetime coordinates, whereas the Latin indices 
shall represent the spatial coordinates, except for $k$ which shall be reserved 
for denoting the wavenumber. 
Lastly, an overprime shall denote differentiation with respect to the conformal 
time coordinate.


\section{Non-minimally coupled, helical electromagnetic fields, quantization 
and power spectrum}

We shall consider the background to be the spatially flat, 
Friedmann-Lema\^itre-Robertson-Walker (FLRW) metric that is described by 
the line-element
\begin{equation}
\d s^2 = a^2(\eta)\, 
\l(-\d\eta^2+\delta_{ij}\, \d x^i\,\d x^j\r),
\end{equation}
where $a(\eta)$ is the scale factor and $\eta$ denotes the conformal time 
coordinate.
We consider the action~\cite{Caprini:2014mja}
\begin{equation}\label{eq:em-action}
S_{\rm em}[A^{\mu},\phi]
=-\frac{1}{16\,\pi} \int \d^{4}x\, \sqrt{-g}\left[J^2(\phi)\, F_{\mu\nu}F^{\mu\nu}
-\frac{\gamma}{2}\, I^2(\phi)\, F_{\mu\nu}{\tilde F^{\mu\nu}}\right],
\end{equation}
where the electromagnetic field tensor is given by 
\begin{equation}
F_{\mu\nu} = \partial_\mu\,A_\nu - \partial_\nu\,A_\mu. 
\end{equation}
The dual of the electromagnetic tensor ${\tilde F^{\mu\nu}}$ is defined as 
\begin{equation}
{\tilde F^{\mu\nu}} = \epsilon^{\mu\nu\alpha\beta}\,F_{\alpha\beta},
\end{equation}
with $\epsilon^{\mu\nu\alpha\beta} = \left(1/\sqrt{-g}\right)\, 
{\cal A}^{\mu\nu\alpha\beta}$, where ${\cal A}^{\mu\nu\alpha\beta}$ 
is the totally antisymmetric Levi-Civita tensor and ${\cal A}^{0123} = 1$.
Clearly, the function $J(\phi)$ describes the non-minimal coupling, while 
the function $I(\phi)$ (along with the parameter $\gamma$) leads to parity 
violation.

\par

We shall assume that there is no homogeneous component to the electromagnetic 
field.
We shall choose to work in the Coulomb gauge wherein $A_0 = 0$ and 
$\partial_i\,A^i = 0$.
In such a gauge, at the quadratic order in the inhomogeneous modes, the 
action describing the electromagnetic field is found to be
\begin{equation}
S[A_i] 
= \frac{1}{4\,\pi}\,\int\,{\rm d}\,\eta\,\int\,{\rm d}^3\,\bm{x}
\left\{J^2\left(\phi\right)
\left[\frac{1}{2}\,A_i^{\prime\,2} - \frac{1}{4}\left(\partial_i\,A_j 
- \partial_j\,A_i\right)^2\right]
+ \gamma\,I^2\left(\phi\right)\,\epsilon^{ijk}\,A_i^\prime\,\partial_j\,
A_k\right\},
\end{equation}
where $\epsilon^{ijk}$ is the three-dimensional completely anti-symmetric 
tensor.
We can vary this action to arrive at the following equation of motion for 
the electromagnetic vector potential:
\begin{equation}
A_i^{\prime\prime} + 2\,\frac{J^\prime}{J}\,A_i^\prime - \nabla^2\,A_i 
= -\frac{\gamma}{J^2}\,\frac{{\rm d}I\,^2}{{\rm d}\,\eta}\,
\delta_{il}\,\epsilon^{lnm}\,\partial_n\,A_m.
\end{equation}

\par

For each comoving wave vector $\bm{k}$, we can define the right-handed 
orthonormal basis $(\varepsilon_1^{\bm{k}}, \varepsilon_2^{\bm{k}}, 
\hat{\bm{k}})$, where
\begin{equation}
\vert\varepsilon^{\bm{k}}_i\vert^2 = 1,\quad
\varepsilon^{\bm{k}}_1\times\varepsilon^{\bm{k}}_2 
= \hat{\bm{k}}\quad{\rm and}\quad
\varepsilon^{\bm{k}}_1\cdot\varepsilon^{\bm{k}}_2
=\hat{\bm{k}}\cdot\varepsilon^{\bm{k}}_{1}  
= \hat{\bm{k}}\cdot\varepsilon^{\bm{k}}_{2}=0.
\end{equation}
While this is a suitable basis for expressing the non-helical modes, 
it is not ideally suited for the helical case as the two helical 
modes would be coupled in this basis.
In such a situation, it is convenient to identify two transverse 
directions to form the helicity basis, wherein the modes decouple, 
as follows~\cite{Durrer:2010mq,Jain:2012jy}:
\begin{equation}
\varepsilon^{\bm{k}}_{\pm} 
= \frac{1}{\sqrt{2}}\left(\varepsilon^{\bm{k}}_1 \pm i\,
\varepsilon^{\bm{k}}_2\right).
\end{equation}
In such a basis, on quantization, the vector potential $\hat{A_i}$ 
can be Fourier decomposed as~\cite{Caprini:2014mja}:
\begin{equation}\label{eq:fourier-decomposition}
{\hat A}_i(\eta,{\bm x}) 
=\sqrt{4\,\pi}\int\frac{\d^3\,{\bm k}}{(2\,\pi)^{3/2}}\,
\sum_{\sigma=\pm}\,\l[\varepsilon^{\bm{k}}_{\sigma i}\,
\hat{b}_{\bm k}^\sigma\, 
{\bar A}_k^\sigma(\eta)\, {\rm e}^{i\,{\bm k}\cdot{\bm x}}
+ \varepsilon^{\bm{k}\,\ast}_{\sigma i}\,
\hat{b}_{\bm k}^{\sigma\dagger}\, {\bar A}_k^\sigma(\eta)\,
{\rm e}^{-i\,{\bm k}\cdot{\bm x}}\r],
\end{equation}
where the Fourier modes ${\bar A}_k^\sigma$ satisfy the differential 
equation
\begin{equation}
{\bar A}_k^{\sigma\,\prime\prime} 
+ 2\,\frac{J^\prime}{J}\, {\bar A}_k^{\sigma\,\prime} 
+ \l(k^2+ \f{\sigma\,\gamma\,k}{J^2}\,\frac{{\rm d}I\,^2}{{\rm d}\,\eta}\r)\,
{\bar A}_k^\sigma = 0.\label{eq:Asigma-diff}
\end{equation}
Note that $\sigma=\pm 1$, and this causes considerable difference 
in the evolution of the modes.
As we shall see, one of the modes will be strongly amplified on 
super-Hubble scales and the extent of amplification will depend 
on the quantity $\gamma$.
The operators $\hat{b}_{\bm k}^{\sigma}$ and $\hat{b}_{\bm k}^{\sigma\dagger}$ 
are the annihilation and creation operators satisfying the following standard 
commutation relations:
\begin{equation}
[{\hat b}_{\bm k}^\sigma,\hat{b}_{\bm k'}^{\sigma'}] 
= [{\hat b}_{\bm k}^{\sigma\dagger},
{\hat b}_{\bm k'}^{\sigma^\prime\dagger}] = 0\quad{\rm and}\quad
[\hat{b}_{\bm k}^\sigma,\hat{b}_{\bm k'}^{\sigma'\dagger}] 
=\delta^{(3)}({\bm k} - {\bm k'})\,\delta_{\sigma\sigma'}.
\end{equation}
Let us define $\cA_k^\sigma = J\,{\bar A}_k^\sigma$. 
In terms of the new variable $\cA_k^\sigma$, 
Eq.~(\ref{eq:Asigma-diff}) can be rewritten as
\begin{equation}
\cA_k^{\sigma\,\prime\prime} + \l(k^2 - \frac{J^{\prime\prime}}{J} 
+ \f{\sigma\,\gamma\,k}{J^2}\,\f{{\rm d}I\,^2}{{\rm d}\,\eta}\r)\,
\cA_k^\sigma 
= 0.\label{eq:cA-eqn-gen}
\end{equation}
In this work, we shall restrict ourselves to the simplest scenarios 
wherein $I = J$. 
In such a case, the above equation simplifies to
\begin{equation}
\cA_k^{\sigma\,\prime\prime}
+ \left(k^2 - \frac{J^{\prime\prime}}{J} 
+ \frac{2\,\sigma\,\gamma\,k\,J^\prime}{J}\right)\cA_k^\sigma = 0.
\label{eq:cA-eqn}
\end{equation}

\par

Since we shall be only interested in the behavior of the helical magnetic 
fields, we shall not discuss the electric field in this work.
Let $\hat{\rho}_{_{\rm B}}$ denote the operator corresponding to the energy 
density associated with the magnetic field.
Upon using the decomposition~(\ref{eq:fourier-decomposition}) of the vector 
potential, the expectation value of the energy density $\hat{\rho}_{_{\rm B}}$ 
can be evaluated in the vacuum state, say, $\vert 0\rangle$, that is 
annihilated by the operator $\hat{b}_{\bm{k}}^\sigma$ for all $\vk$ and 
$\sigma$.
It can be shown that the spectral energy density of the magnetic field can 
be expressed in terms of the modes ${\bar A}_k^\sigma$ and the coupling 
function $J$ as follows~\cite{Durrer:2010mq,Anber:2006xt,Caprini:2014mja}:
\begin{equation}
\psb(k)
=\frac{\d\langle 0 \vert \hat{\rho}_{_{\rm B}}\vert 0 \rangle}
{\d\,{\rm ln}\,k}
= \frac{J^2(\eta)}{4\,\pi^{2}}\,\frac{k^{5}}{a^{4}(\eta)}
\left[\lvert {\bar A}_k^{+}(\eta)\rvert^2 
+ \lvert {\bar A}_k^{-}(\eta)\rvert^2\right].\label{eq:psb}
\end{equation}
The spectral energy density $\psb(k)$ is referred to as 
the power spectrum for the magnetic field. 
A flat or scale invariant magnetic field spectrum corresponds to a constant, 
\ie $k$-independent, ${\mathcal P}_{_{\rm B}}(k)$.


\section{Power spectra of the helical magnetic fields generated in de Sitter 
inflation}\label{sec:psbi}

Let us consider the simple case of de Sitter inflation, wherein the 
scale factor is given by
\begin{equation}
a(\eta) = -\frac{1}{H_0\,\eta},\label{eq:sf-ds}
\end{equation}
where $H_0$ is the value of the Hubble parameter during inflation.
In order to solve for the electromagnetic modes, we need to choose a form 
of the coupling function.
In keeping with the expressions for the coupling functions that have been 
adopted in earlier work~\cite{Caldwell:2011ra,Jain:2012ga,Jain:2012vm}, we 
shall work with a coupling function that can be written as a simple power 
of the scale factor as follows:
\begin{equation}
J(\eta) = J_0\,[a(\eta)]^n
=\f{J_0}{(-H_0\,\eta)^n}.\label{eq:Jh}
\end{equation}
We shall set $J_0=(-H_0\,\ee)^n$, where $\ee$ denotes the conformal time at
the end of inflation. 
This choice ensures that $J$ reduces to unity at $\ee$.

\par

For the form of the coupling function given by Eq.~(\ref{eq:Jh}), 
the solutions to the electromagnetic modes satisfying  Eq.~(\ref{eq:cA-eqn}) 
can be written as follows~\cite{NIST:DLMF,Gradshteyn:2007,Caprini:2014mja}:
\begin{equation}
\cA_k^\sigma\left(\eta\right) 
= \f{1}{\sqrt{2\,k}}
\l[G_{n}(\sigma\,\xi, -k\,\eta) 
+ i\,F_{n}(\sigma\,\xi, -k\,\eta)\r],\label{eq:hm}
\end{equation}
where $G_{n}(\sigma\,\xi, -k\,\eta)$ and $F_{n}(\sigma\,\xi, -k\,\eta)$ 
represent the irregular and regular Coulomb functions respectively and $\xi 
= -n\,\gamma$.
For $- k\,\eta \ll \sigma\,\xi$, which corresponds to modes of 
interest, the contribution of $F_{n}(\sigma\,\xi, -k\,\eta)$ to 
the mode is negligible. 
We also find that the mode with negative helicity (\ie with $\sigma=-1$) 
is amplified in comparison to the positive helicity mode. 
In this regime, the irregular Coulomb function can be written in terms 
of the modified Bessel function $K_\nu(z)$ as follows~\cite{NIST:DLMF}:
\begin{equation}
G_L\l(y,z\r)
= \f{2\,(2\,y)^L}{(2\,L+1)!\,C_L(y)}\,
\l(2\,y\,z\r)^{1/2}\,K_{2\,L+1}\l(\sqrt{8\,y\,z}\r),
\end{equation}
where $C_L(y)$ is given by
\begin{equation}\label{eq:Cl}
C_L(y) = \f{2^L\,{\rm e}^{-\pi\, y/2}\,
\l\vert\Gamma\l(L+1+i\, y\r)\r\vert}{\Gamma\l(2\,L + 2\r)}.
\end{equation}
Hence the modes~(\ref{eq:hm}) reduce to
\begin{equation}
\cA_k^{-}(\eta) 
\simeq \sqrt{-\f{2\eta}{\pi}}\,
{\rm e}^{-\pi\xi}\,K_{2n+1}\l(\sqrt{8\,\xi\, k\,\eta}\r).
\end{equation}
Also, for $-k\,\eta \ll 1/\xi$, which corresponds to late times during 
inflation, using the properties of the modified Bessel function for 
small arguments, we obtain that~\cite{Abramowitz:2001}
\begin{equation}\label{eq:cA-limit}
\cA_k^{-}(\eta) 
\simeq \sqrt{-\f{\eta}{2\pi}}\,{\rm e}^{-\pi\xi}\,
\Gamma\l(\vert 2n+1\vert\r)\,
\vert 2\,\xi\, k\,\eta\vert^{-\vert n+\frac{1}{2}\vert}.
\end{equation}
Therefore, using Eq.~(\ref{eq:psb}), the power spectrum for the 
magnetic field evaluated as $\ee\to 0$ can be expressed as~\cite{Caprini:2014mja}
\begin{equation}
\psb(k) 
\simeq \frac{H_0^4}{8\,\pi^3}\,{\rm e}^{-2\,\pi\,\xi}\,
\l[\Gamma(\vert 2\,n+1\vert)\r]^2\,
\l\vert 2\,\xi \r\vert^{-2\,\vert n+\f{1}{2}\vert}\,
(k\,\ee)^{5-2\vert n+\frac{1}{2}\vert}.
\end{equation}
Evidently, the spectral index $n_{_{\rm B}}$ of the power spectrum of 
the magnetic field is 
given by
\begin{equation}
n_{_{\rm B}} = 5 - 2\,\l\vert n+\frac{1}{2}\r\vert.
\end{equation}
This implies that we obtain a scale invariant power spectrum for the 
magnetic field when $n=-3$ or $n=2$, just as in the non-helical case.
Note that the power spectrum is completely independent of time in these 
situations.
However, when $n<0$, it is found that the energy in the electromagnetic
field grows rapidly at late times.
Such a growth leads to severe backreaction and can result in the 
termination of inflation within a few e-folds~\cite{Caldwell:2011ra,
Jain:2012ga,Jain:2012vm}.
Because of this reason, in this work, we shall focus only on the cases
wherein $n>0$.


\section{Formal structure of the three-point function}\label{sec:cc}

In the preceding section, when we had considered the evolution of the 
electromagnetic modes, for simplicity, we had assumed the non-minimal 
coupling $J(\eta)$ to be given by Eq.~(\ref{eq:Jh}).
However, in order to evaluate the cross-correlation between the perturbation
in the scalar field and the magnetic field, other than $J$, we shall also 
require the function $J_\phi=\d J/\d \phi$.
Since
\begin{equation}
J_\phi=\f{J'}{\phi'},
\end{equation}
and, as $J(\eta)$ has been chosen already [cf. Eq.~(\ref{eq:Jh})], clearly, 
we can arrive at $\d J/\d \phi$ if we know $\phi'$.
This can be achieved by choosing a potential $V(\phi)$ to drive 
the scalar field.
Then, a suitable $J(\phi)$ can lead to the desired $J(\eta)$.

\par

Let $\phi$ be an auxiliary scalar field described by the potential
$V(\phi)$ that is evolving in de Sitter spacetime described by the 
scale factor~(\ref{eq:sf-ds}).
In such a situation, the homogeneous scalar field $\phi$ satisfies 
the following equation of motion:
\begin{equation}
\phi'' - \f{2}{\eta}\,\phi' + a^2\,V_\phi= 0,
\end{equation}
where $V_\phi=\d V/\d\phi$.
If we now assume that $V(\phi) = -3\,n\,\Mp\,H_0^2\,\phi$~\cite{Caldwell:2011ra}, 
where $n$ is a constant, then it is straightforward to show that, for a suitable 
choice of initial conditions, the solution to the above equation governing the
scalar field can be written as
\begin{equation}
\phi(\eta) 
=-n\,\Mp\,{\rm ln}\, \eta.
\end{equation}
Hence, upon setting $J(\phi)=J_0\, \exp\,(\phi/\Mpl)$, we can arrive 
at the desired behavior of $J(\eta)$ [as given by Eq.~(\ref{eq:Jh})]
that we had worked with.  
With $J(\phi)$ at hand, we can, evidently, obtain $J_\phi$ to be
\begin{eqnarray}
J_\phi= \frac{J(\phi)}{\Mp},
\end{eqnarray}
thereby arriving at the required quantities related to the background.

\par

Note that, we have assumed the electromagnetic field to be inhomogeneous. 
Therefore, in order to calculate the three-point function involving 
perturbations in the scalar field and the electromagnetic vector 
potential, we need the interaction Hamiltonian at the third order 
in the perturbations. 
This can be arrived at by perturbing the electromagnetic 
action~(\ref{eq:em-action}) with respect to the scalar field. 
It is straightforward to show that the third order action for the case 
$I=J$ is given by
\begin{equation}
S[A_i] 
= \frac{1}{2\,\pi}\,\int\,{\rm d}\,\eta\,\int\,{\rm d}^3\,\bm{x}\,
J\,J_\phi\,\delta\phi\,
\l\{\l[\f{1}{2}\,A_i^{\prime\,2} 
- \f{1}{4}\,\l(\pa_i\,A_j - \pa_j\,A_i\r)^2\r]
+ \gamma\,\epsilon^{ijk}\,A_i^\prime\,\partial_j\,A_k\r\}.
\end{equation}
The interaction Hamiltonian can be obtained from this third order action. 
It is found to be
\begin{equation}
H_{\rm int} = \f{1}{4\,\pi}\,\int{\rm d}^3\,\bm{x}\,
J\,J_\phi\,\delta\phi\,\l\{A_i^{\prime\,2} 
+ \f{1}{2}\,\l(\partial_i\,A_j - \partial_j\,A_i\r)^2\r\}.
\label{eq:Hinth} 
\end{equation}
Two points need to be stressed regarding this interaction 
Hamiltonian~\cite{Bartolo:2015dga,Bartolo:2014hwa}. 
Firstly, the parity violating term does not contribute to the Hamiltonian. 
This implies that the formal structure of the resulting three-point function 
will be largely similar to the non-helical case that has been considered 
earlier~\cite{Caldwell:2011ra,Motta:2012rn,Jain:2012ga,Jain:2012vm}.
Secondly, as we shall see below, the effects of non-zero helicity will be
essentially encoded in the way it affects the evolution of the modes.

\par

The cross-correlation between the perturbation in the scalar field and the 
magnetic field in real space is defined as
\begin{eqnarray}
& &\!\!\!\!\!\!\!\!\!\!\!\!\!\!\!\!\!\!\!\!\!\!\!\!
\l\langle\frac{\hat{\delta\phi}(\eta,\bm{x})}{\Mpl}\,
\hat{B}^i(\eta,\bm{x})\,\hat{B}_i(\eta,\bm{x})\r\rangle\nn\\ 
&=& \int \f{\d^3{\bm k}_1}{(2\,\pi)^{3/2}}
\int \f{\d^3{\bm k}_2}{(2\,\pi)^{3/2}}
\int \f{\d^3{\bm k}_3}{(2\,\pi)^{3/2}}\,
\l\langle \f{\hat{\delta\phi}_{\bm{k_1}}\!(\eta)}{\Mpl}\,
\hat{B}^i_{\bm{k_2}}(\eta)\,\hat{B}_{i\,\bm{k_3}}(\eta)\r\rangle\,
{\rm e}^{i\,({\bm k_1}+{\bm k_2}+{\bm k_3})\cdot{\bm x}},
\end{eqnarray}
where the components $B_i$ of the magnetic field are related to the 
vector potential $A_i$ through the relation
\begin{equation}
B_i = \f{1}{a}\,\epsilon_{ijl}\,\partial_j\,A_l,
\end{equation}
while $\delta\phi_{\bm k}$ and  $B^i_{\bm k}$ denote the Fourier modes 
associated with the perturbation in the scalar field and the $i$-th 
component of the magnetic field.
As per the standard rules of perturbative quantum field theory, the 
cross-correlation between the perturbation in the scalar field and the 
magnetic field in Fourier space, evaluated at the end of inflation, is
given by~\cite{Caldwell:2011ra,Jain:2012vm}
\begin{equation}
\l\langle \frac{{\hat{\delta\phi}_{\bm{k_1}}\!(\ee)}}{\Mpl}\,
\hat{B}^i_{\bm{k_2}}(\eta_{\rm e})\,
\hat{B}_{i\,\bm{k_3}}(\eta_{\rm e})\r\rangle
= -i\, \int_{\eta_{\rm i}}^{\eta_{\rm e}} {\rm d}\eta\,
\l\langle\l[\frac{\hat{\delta\phi}_{\bm{k_1}}}{\Mpl}(\eta_{\rm e})\,
\hat{B}^i_{\bm{k_2}}(\eta_{\rm e})
\,\hat{B}_{i\,\bm{k_3}}(\eta_{\rm e}),\hat{H}_{\rm int}(\eta)\r]\r\rangle,
\label{eq:tpffnh}
\end{equation}
where ${\hat H}_{\rm int}$ is the operator associated with the 
Hamiltonian~(\ref{eq:Hinth}) and the square brackets indicates 
the commutator.

\par

We had discussed the quantization of the helical electromagnetic field
in the previous section.
At this stage, let us discuss the quantization of the perturbation in 
the scalar field~$\delta\phi$.
The perturbation in the scalar field can be quantized in terms of the 
corresponding Fourier modes, say, $f_k$, as 
\begin{equation}
\hat{\delta\phi}\l(\eta,{\bm x}\r) 
=\int\frac{\d^3\,{\bm k}}{(2\,\pi)^{3/2}}\,
\l[\hat{a}_{\bm k}\, f_k(\eta)\,{\rm e}^{i\,{\bm k}\cdot{\bm x}}
+ \hat{a}_{\bm k}^{\dagger}\, f_k^{\ast}(\eta)\,
{\rm e}^{-i\,{\bm k}\cdot{\bm x}}\r],
\end{equation}
where the annihilation and creation operators $\hat{a}_{\bm k}$ and 
$\hat{a}_{\bm k}^\dag$ satisfy the following standard commutation
relations:
\begin{equation}
[\hat{a}_{\bm k},\hat{a}_{\bm k'}] 
= [\hat{a}_{\bm k}^{\dagger},\hat{a}_{\bm k'}^{\dagger}] = 0,\quad
[\hat{a}_{\bm k},\hat{a}_{\bm k'}^{\dagger}] 
=\delta^{(3)}({\bm k} - {\bm k'}),
\end{equation}
and the Fourier modes $f_k$ satisfy the differential equation
\begin{equation}
f_k'' + 2\,\cH\,f_k' + k^2\,f_k=0,\label{eq:fk-diff}
\end{equation}
with $\cH=a^\prime/a$ being the conformal Hubble parameter.
It is useful to point out here that, since the potential $V(\phi)$ 
is linear in the field, the potential does not directly influence 
the evolution of the perturbation in the scalar field.
The scalar modes $f_k$ are determined only by the behavior of the 
scale factor.

\par

Let us now define
\begin{eqnarray}
\l\langle \f{\hat{\delta\phi}_{\bm{k_1}}(\ee)}{\Mpl}
\hat{B}^i_{\bm{k_2}}(\ee)\,
\hat{B}_{i\,\bm{k_3}}(\ee)\r\rangle
\equiv (2\,\pi)^{-3/2}\,
G_{\delta\phi B B}\l(\bm{k_1},\bm{k_2},\bm{k_3}\r)\,
\delta^{(3)}\l(\bm{k_1}+\bm{k_2}+\bm{k_3}\r),\nn\\
\label{eq:cch}
\end{eqnarray}
where $G_{\delta\phi B B}(\bm{k_1},\bm{k_2},\bm{k_3})$ is the 
cross-correlation of our interest in Fourier space.
We can now determine the three-point function using the 
expression~(\ref{eq:tpffnh}), along with the form of the interaction 
Hamiltonian~(\ref{eq:Hinth}) and Wick's theorem that applies to the 
products of Gaussian operators.
We find that the cross-correlation 
$G_{\delta\phi B B}(\bm{k_1},\bm{k_2},\bm{k_3})$
can be expressed as
\begin{eqnarray}
G_{\delta\phi B B}(\bm{k_1},\bm{k_2},\bm{k_3}) 
&=& \f{8\,\pi}{\Mp\,a^2(\ee)}\,
\l[(\vkb \cdot \vkc)\,\delta_{qn} - k_{2n}\,k_{3q}\r]\,
\sum_{\sigma\,\sigma'}\,f_{\ka}(\ee)\,\bA^\sigma_{\kb}(\ee)\,
\bA^{\sigma^\prime}_{\kc}(\ee)\nn\\
& &\times\, \biggl\{\varepsilon_{\sigma q}(\vkb)\,
\varepsilon^\ast_{\sigma l}(\vkb)\,
\varepsilon_{\sigma'n}(\vkc)\,\varepsilon^\ast_{\sigma'l}(\vkc)\, 
\cG_1^{\sigma\sigma^\prime}(\bm{k_1},\bm{k_2},\bm{k_3}) \nn\\
& &-\, \biggl[\frac{(\vkb \cdot \vkc)}{k_2\,k_3}\,\varepsilon_{\sigma q}(\vkb)\,
\varepsilon^\ast_{\sigma r}(\vkb)\,
\varepsilon_{\sigma'n}(\vkc)\,
\varepsilon^\ast_{\sigma^\prime r}(\vkc) \nn\\
& &-\, \frac{\l(k_{2l}\,k_{3r}\r)}{k_2\,k_3}\,\varepsilon_{\sigma q}(\vkb)\,
\varepsilon^\ast_{\sigma r}(\vkb)\,
\varepsilon_{\sigma'n}(\vkc)\,
\varepsilon^\ast_{\sigma'l}(\vkc)\biggr]\,
\cG_2^{\sigma\sigma^\prime}(\bm{k_1},\bm{k_2},\bm{k_3})\biggr\} \nn\\
&& +\,{\rm complex~conjugate},
\end{eqnarray}
where the quantities $\cG_1^{\sigma\sigma'}(\bm{k_1},\bm{k_2},\bm{k_3})$
and $\cG_2^{\sigma\sigma'}(\bm{k_1},\bm{k_2},\bm{k_3})$ are described
by the integrals
\begin{subequations}\label{eq:int-h-eta}
\begin{eqnarray}
\cG_1^{\sigma\sigma'}(\bm{k_1},\bm{k_2},\bm{k_3}) 
&=& i\,\int_{\eta_{\rm i}}^{\eta_{\rm e}} {\rm d}\eta\,
J\,\frac{{\rm d}J}{{\rm d}\phi}\,f_{\ka}^\ast(\eta)\,
\bA_{\kb}^{\prime\ast\sigma}(\eta)\,
{\bar A}_{\kc}^{\prime\ast\sigma^{\prime}}(\eta),\label{eq:int-h1-eta}\\
\cG_2^{\sigma\sigma'}(\bm{k_1},\bm{k_2},\bm{k_3}) 
& =& i\,k_2\,k_3\,\int_{\eta_{\rm i}}^{\eta_{\rm e}} {\rm d}\eta\,
\l(J\,\frac{{\rm d}J}{{\rm d}\phi}\r)\,
f_{\ka}^\ast(\eta)\,{\bar A}_{\kb}^{\ast\sigma}(\eta)\,
{\bar A}_{\kc}^{\ast\sigma^{\prime}}(\eta).
\label{eq:int-h2-eta}
\end{eqnarray}
\end{subequations}


\section{Numerical evaluation of the three-point function}\label{sec:ne}

It is evident that evaluating the cross-correlation of our interest 
involves integrals over products of the electromagnetic modes and 
the modes corresponding to the perturbation in the scalar field.
Earlier, we had solved for the electromagnetic modes analytically 
in order to arrive at the power spectrum.
The analytical solutions entail writing the modes in terms of Coulomb 
functions [cf. Eq.~(\ref{eq:hm})], which seem non-trivial to integrate.
Therefore, in order to evaluate the three-point function, we shall 
resort to numerical computations.
In order to obtain the three-point function, we need to solve 
for the electromagnetic modes as well as for the modes of the scalar 
perturbations numerically. 
Thereafter, we need to integrate these modes in order to arrive at the 
complete three-point function.


\subsection{Evolution of the modes}

Let us first discuss the method we shall adopt to numerically solve
for the electromagnetic and scalar modes $\bA_k^\sigma$ and $f_k$.
The most efficient time variable to perform the numerical analyses in 
inflationary scenarios is the e-fold~$N$.
In terms of e-folds, Eq.~(\ref{eq:Asigma-diff}) can be written as
\begin{equation}\label{eq:Asigma-diff-N-full}
\frac{\d^2 \bA_k^\sigma}{\d N^2} + \l(\frac{\cH_N}{\cH}
+ 2\,\frac{J_N}{J}\r)\frac{\d \bA_k^\sigma}{\d N} 
+ \l(\frac{k^2}{\cH^2} + \frac{2\,\sigma\,\gamma\,k}{\cH}\,
\frac{J_N}{J}\r)\,\bA_k^\sigma = 0,
\end{equation}
where the subscript $N$ refers to a derivative with respect to the 
e-fold.
For the case of de Sitter inflation and the form of the coupling function 
under consideration, we obtain that $\cH_N/\cH=1$ and $J_N/J=n$.
Under these conditions, the above differential equation simplifies to
\begin{equation}
\frac{\d^2 \bA_k^\sigma}{\d N^2} + \l(2\,n + 1\r)
\frac{\d \bA_k^\sigma}{\d N} + \l(\frac{k^2}{\cH^2} 
+ \frac{2\,\sigma\,\gamma\,k\,n}{\cH}\r)\,\bA_k^\sigma = 0.
\label{eq:Asigma-diff-N}
\end{equation}
Similarly, Eq.~(\ref{eq:fk-diff}) governing the evolution of the scalar 
modes can be rewritten as
\begin{equation}\label{eq:fk-diff-N-full}
\frac{\d^2 f_k}{\d N^2} 
+ \l(2 + \frac{\cH_N}{\cH}\r)\frac{\d f_k}{\d N} 
+ \frac{k^2}{\cH^2}\,f_k = 0,
\end{equation}
which, for the case of de Sitter inflation, simplifies to
\begin{equation}
\frac{\d^2 f_k}{\d N^2} + 3\,\frac{\d f_k}{\d N} 
+ \frac{k^2}{\cH^2}\,f_k = 0.\label{eq:fk-diff-N}
\end{equation}

\par

Recall that, during inflation, in the case of the scalar and tensor 
modes, the standard Bunch-Davies initial conditions are imposed on 
the modes when they are well inside the Hubble radius. 
It is clear from Eq.~(\ref{eq:cA-eqn}) governing the dynamics of the
electromagnetic modes that, at very early times, it is the term 
involving $k^2$ that dominates the other two terms within the 
parentheses. 
In fact, we find that, for the coupling function of our choice 
[cf.~Eq.~(\ref{eq:Jh})], it is the second term that dominates 
the third during the early stages.
These properties permit us to impose Bunch-Davies like initial 
conditions on the modes, and evolve them thereafter.
Numerically, we shall impose the initial conditions when $k=300\,
\sqrt{J''/J}$, corresponding to the e-fold, say, $N_i$.
(This specific choice will be justified in the next section.)
In terms of e-folds, the standard initial conditions on the modes can 
be expressed as follows:
\begin{subequations}
\begin{eqnarray}
\bA_k^{\sigma\,i} &=& \frac{1}{J(N_i)\,\sqrt{2\,k}},\\
\f{\d \bA_k^{\sigma\,i}}{\d N} 
&=& -\frac{n}{J(N_i)\,\sqrt{2\,k}}
-\f{i\,k}{\cH(N_i)\,J(N_i)\,\sqrt{2\,k}}.
\end{eqnarray}
\end{subequations}
We then solve Eq.~(\ref{eq:Asigma-diff-N}) with these initial conditions  
using a fifth order Runge-Kutta routine to obtain the behavior of~$\bA_k^\sigma$.
The differential equation~(\ref{eq:fk-diff-N}) governing the scalar modes 
can be solved for in a similar manner.
The initial conditions on the scalar modes are given by
\begin{subequations}
\begin{eqnarray}
f_k^i &=& \frac{1}{a(N_i)\,\sqrt{2\,k}},\\
\frac{\d f_k^i}{\d N} 
&=& \frac{1}{\cH(N_i)\,a(N_i)\,\sqrt{2\,k}}\l[-i\,k-\cH(N_i)\r].
\end{eqnarray}
\end{subequations}
After imposing these initial conditions on the modes, we shall evolve them 
till about $42$ e-folds for $n=1$ and up to $30$ e-folds for the $n=2$ 
case (the reason for these choices will be explained later).

\par

In Fig.~\ref{fig:bAk}, we have plotted the numerical solutions 
for the helical as well as the non-helical electromagnetic modes
for the cases of $n=1$ and $n=2$.
It is evident from the plots that the modes with negative helicity
(\ie with $\sigma=-1$) are amplified when compared to the non-helical 
case, around the time they leave the Hubble radius.
Also, one finds that the amplitude of the modes with positive
helicity (\ie with $\sigma=1$) are suppressed when compared with
the non-helical modes around Hubble exit.
Moreover, the amplification and suppression is more in the case of 
$n=2$ than $n=1$.
This is to be expected because of the reason that, larger the $n$, 
larger is the amplitude of the parity violating term [cf. 
Eq.~(\ref{eq:Asigma-diff-N})].
\begin{figure}[!h]
\begin{center}
\includegraphics[width=7.50cm]{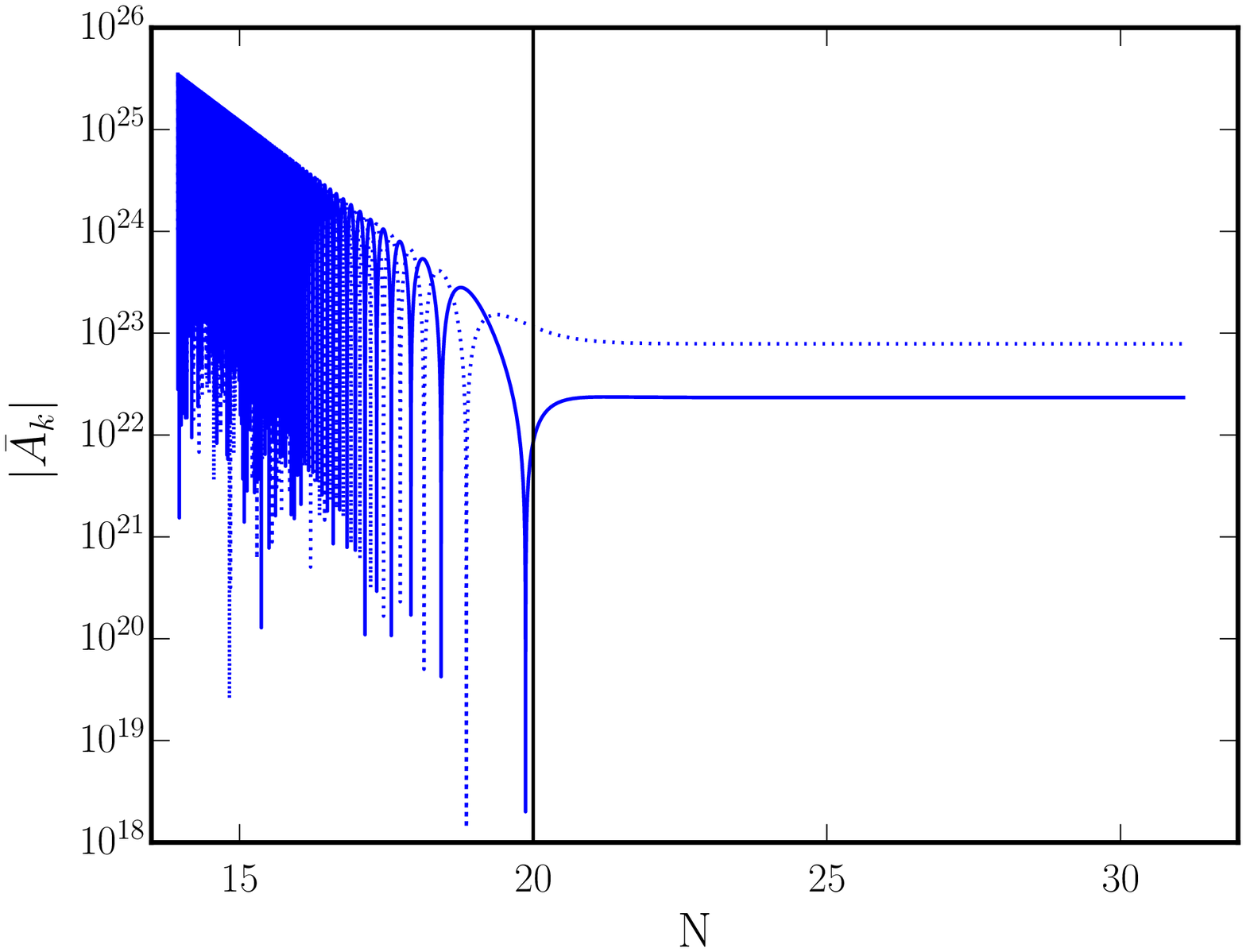}
\includegraphics[width=7.50cm]{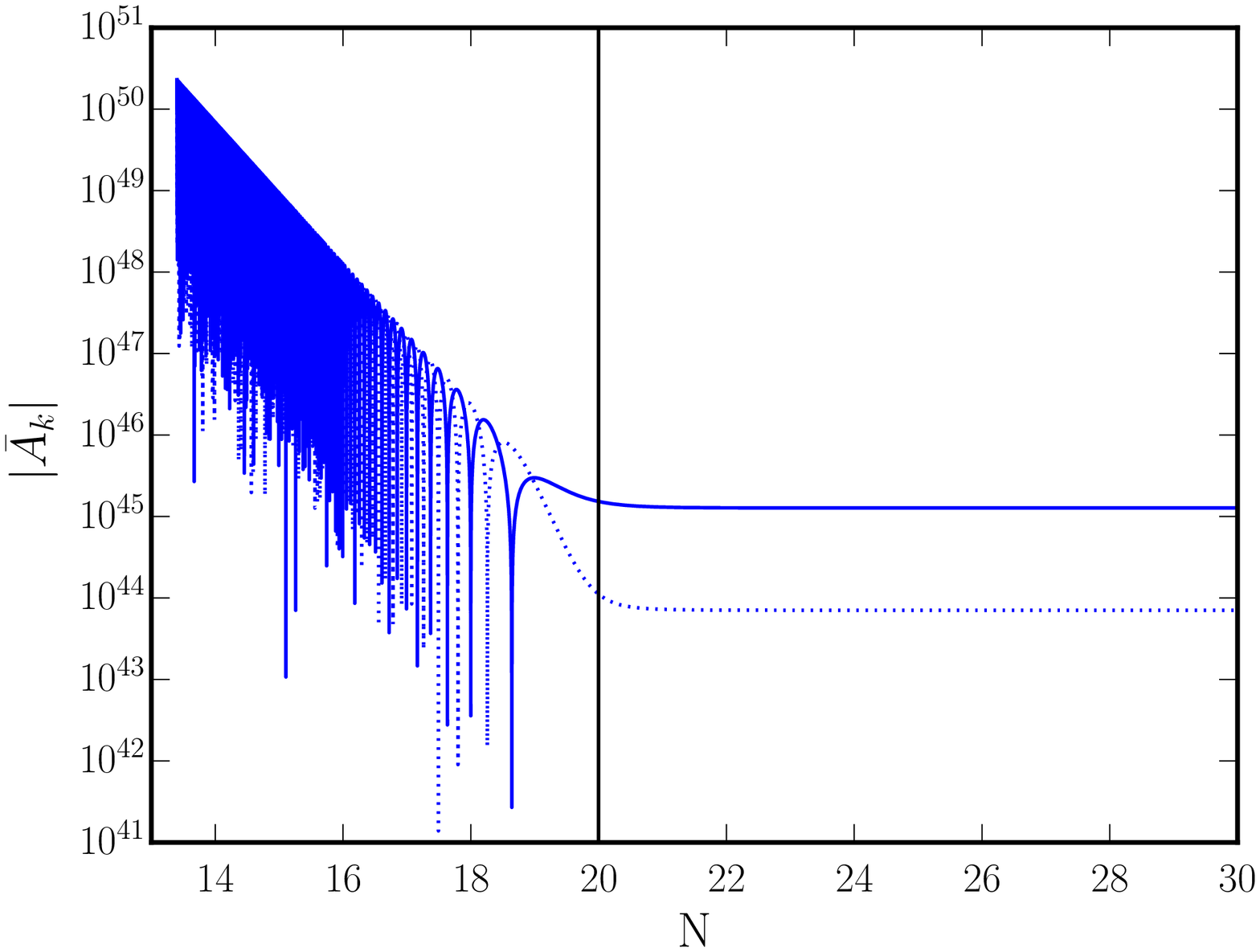}\\
\includegraphics[width=7.50cm]{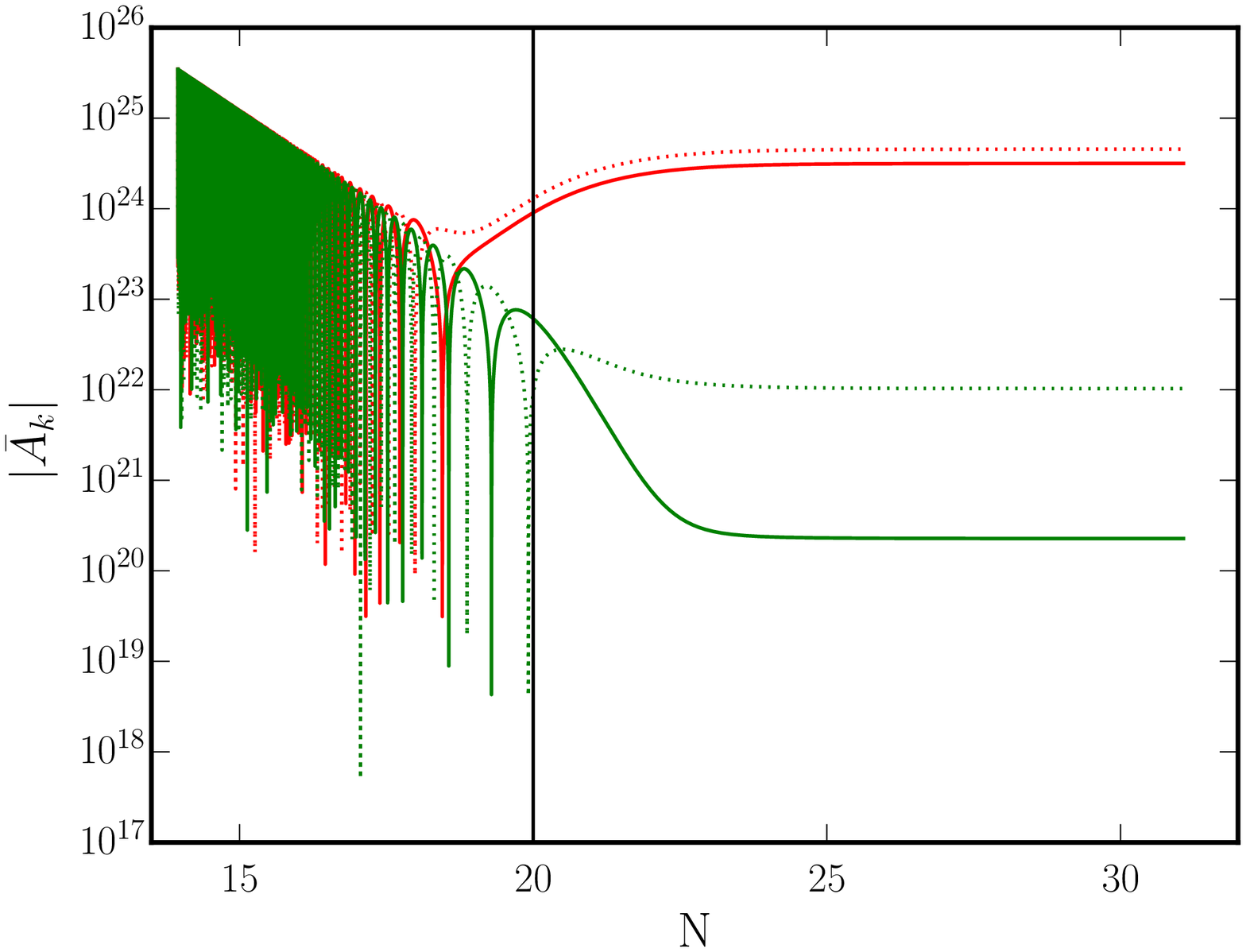}
\includegraphics[width=7.50cm]{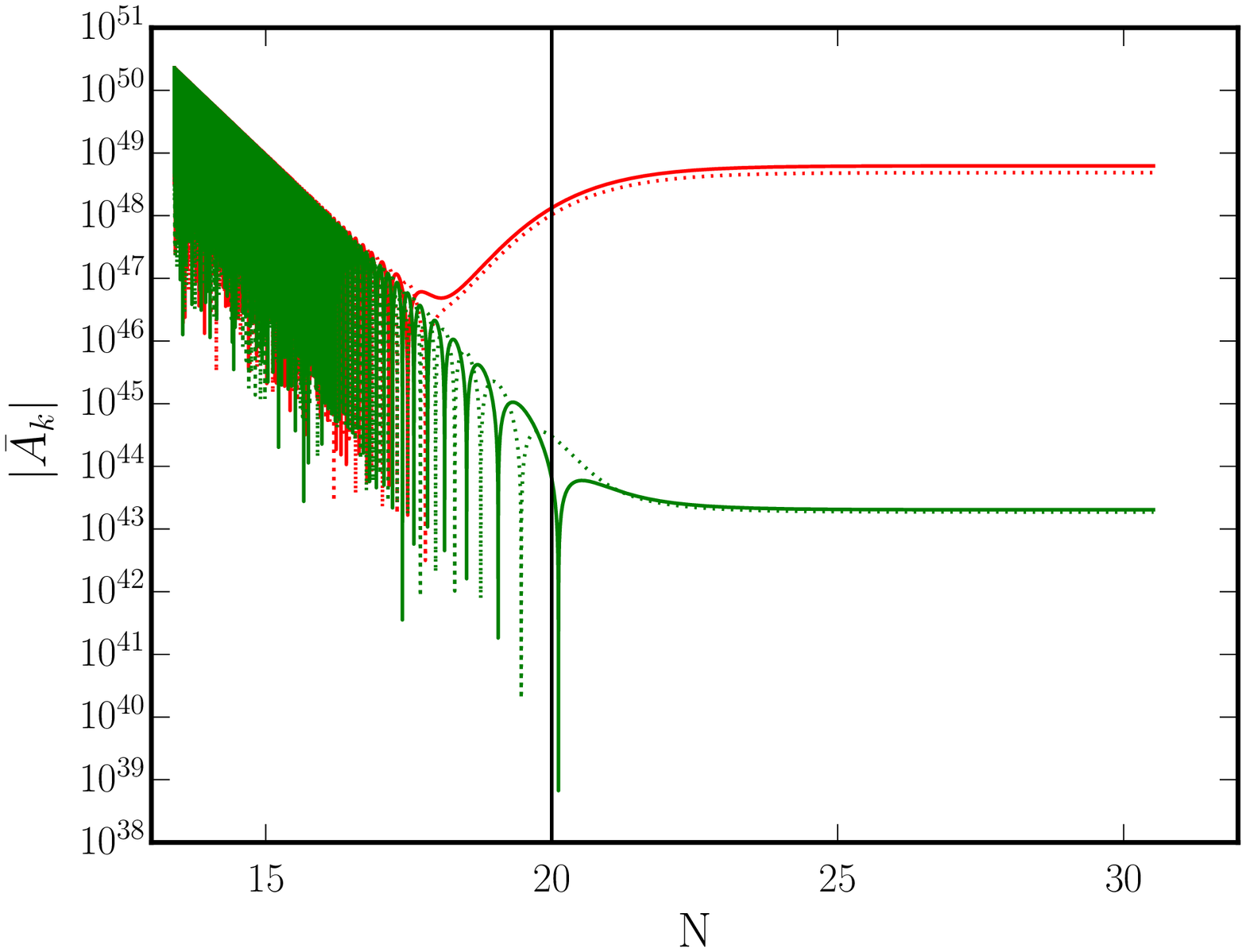}
\end{center}
\caption{The evolution of the mode $\bA_k^\sigma$ has been plotted as a function
of e-folds $N$ for $n=1$ (on the left) and $n=2$ (on the right) for the 
wavenumber $k=0.002\, {\rm Mpc}^{-1}$ for the cases $\gamma=0$ (on top, in 
blue) and $\gamma=2$ (at the bottom, in red and green).
We have plotted the absolute values of the real (solid lines) and 
imaginary (dashed lines) parts separately so that the oscillations 
are visible.
The black vertical lines in each plot indicate the e-fold at which 
the modes leave the Hubble radius.
Note that, around the time when the modes leave the Hubble radius, the 
$\sigma=-1$ mode (in red) is amplified when compared to the non-helical 
case, whereas the $\sigma=1$ mode (in green) is suppressed.
Also, as expected, the amplification and suppression are more in the
$n=2$ case than in the $n=1$ case.}
\end{figure}\label{fig:bAk}

At this stage, we should comment on the amplitude of the parity violating 
parameter~$\gamma$ that we have worked with.
As is well known, magnetic fields possess anisotropic stress.
Such anisotropic stresses can act as a source for the primordial gravitational 
waves and boost their amplitude~\cite{Caprini:2014mja}.
Since the strength of the helical modes are amplified more compared to the 
non-helical modes, one finds that they will in turn affect the amplitude of the 
tensor modes to a larger extent. 
The contribution of the helical magnetic fields to the primordial gravitational waves and the 
current upper bound on the tensor-to-scalar ratio impose
constraints on a combination of the scale of inflation $H/\Mp$ and the parameter 
$\xi$ (recall that, $\xi$ is related to $\gamma$ through the expression
$\xi=-n\,\gamma$).
We find that the effects of these helical fields on the primordial
gravitational waves have been calculated explicitly 
for $n < 0$ (in this context, see Ref.~\cite{Caprini:2014mja}), but not for the 
cases of our interest, \ie $n=1$ and $n=2$.
If we extrapolate the arguments for $n < 0$ to our cases and use the
current upper bound on the tensor-to-scalar ratio, \viz $r \lesssim 10^{-2}$, 
then we find that $\xi \lesssim 10$.
We should add that, even if we vary $r$ or the quantity $H/\Mp$ over a reasonably 
wide range, the upper limit on $\xi$ is not affected substantially.
This is consistent with $\gamma=2$ that we have worked with here. 

\subsection{Evaluation of the three-point function}

Before we go on to consider the cross-correlation of our interest, let
us make a few clarifying remarks concerning the numerical evaluation of 
the inflationary two-point and three-point functions involving the scalar 
and the tensor perturbations.
The inflationary correlation functions are formally expected to be 
evaluated at the end of inflation.
However, it is well known that, during inflation, the amplitude of the 
scalar and the tensor perturbations freeze on super-Hubble scales (apart
from some peculiar situations).
This behavior makes it convenient for the numerical evaluation of the 
power spectra, since they can be evaluated soon after the modes leave 
the Hubble radius. 
Typically, the initial conditions are imposed when the modes are 
sufficiently inside the Hubble radius [say, when $k/(a\,H) \simeq 
10^{2}$] and the power spectra are evaluated when the modes are 
sufficiently outside [say, when $k/(a\,H) \simeq 10^{-5}$].
As far as the three-point functions are concerned, apart from arriving
at the modes, we also need to integrate over them from very early to 
late times.
One can show that, due to the above-mentioned freezing of the amplitude 
of the perturbations, the super-Hubble contributions to the three-point 
functions are insignificant (in this context, see, for instance, 
Refs.~\cite{Hazra:2012yn,Sreenath:2013xra,Sreenath:2014nca}).
Since the three-point function involves an arbitrary triangular
configuration of wavevectors, to arrive at them, the initial
conditions are imposed when the mode with the smallest of the 
wavenumbers is sufficiently inside the Hubble radius (say, at
the e-fold $N_i$) and the integrals involved are evaluated until 
the mode with the largest of the wavenumbers is adequately 
outside (say, at $N_s$)~\cite{Chen:2006xjb,Chen:2008wn,Hazra:2012yn,
Sreenath:2013xra,Sreenath:2014nca}.

\par

There is yet another point to be attended to when evaluating the
three-point functions numerically. 
All the modes will oscillate strongly in the sub-Hubble domain.
Therefore, in order to evaluate the integrals involved, analytically, 
a small parameter, say, $\kappa$, is introduced to achieve an 
exponential cut-off and thereby regulate these 
oscillations~\cite{Maldacena:2002vr}.
Actually, such a regulation is essential to ensure the correct choice 
of the perturbative quantum vacuum~\cite{Maldacena:2002vr,Seery:2005wm}.
One eventually considers the vanishing limit of the parameter $\kappa$ 
to arrive at the final forms of the three-point functions. 
The regulator proves to be very convenient in the numerical efforts,
as it can ensure the convergence of the integrals. 
But, a couple of exercises need to be carried out to identify an apt 
value for the cut-off parameter such that the resulting three-point 
function is ideally independent of the choice of $\kappa$, $N_i$ and 
$N_s$.
We find that the above arguments for the scalar and the tensor perturbations
can be applied to the cross-correlation of our interest for the $n=1$
case.
However, the $n=2$ case poses a peculiar difficulty not often encountered 
in the three-point functions involving the scalars and the tensors, and needs
to handled with care.
We shall make use of the analytical expressions available in the non-helical
case to check the accuracy of our numerical computations.

\par

\begin{figure}[!t]
\begin{center}
\includegraphics[height=6.00cm,width=7.50cm]{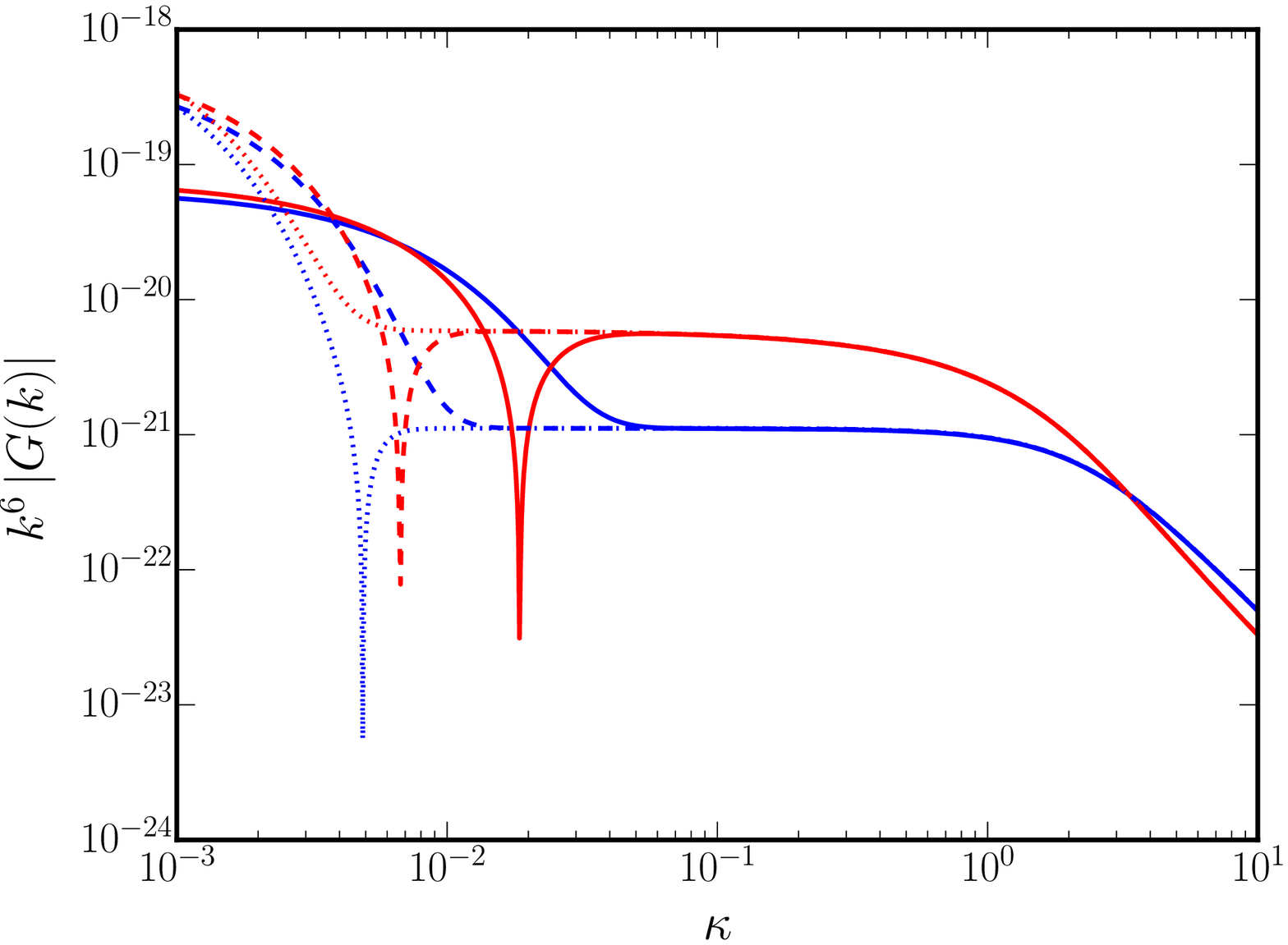}
\includegraphics[height=6.00cm,width=7.50cm]{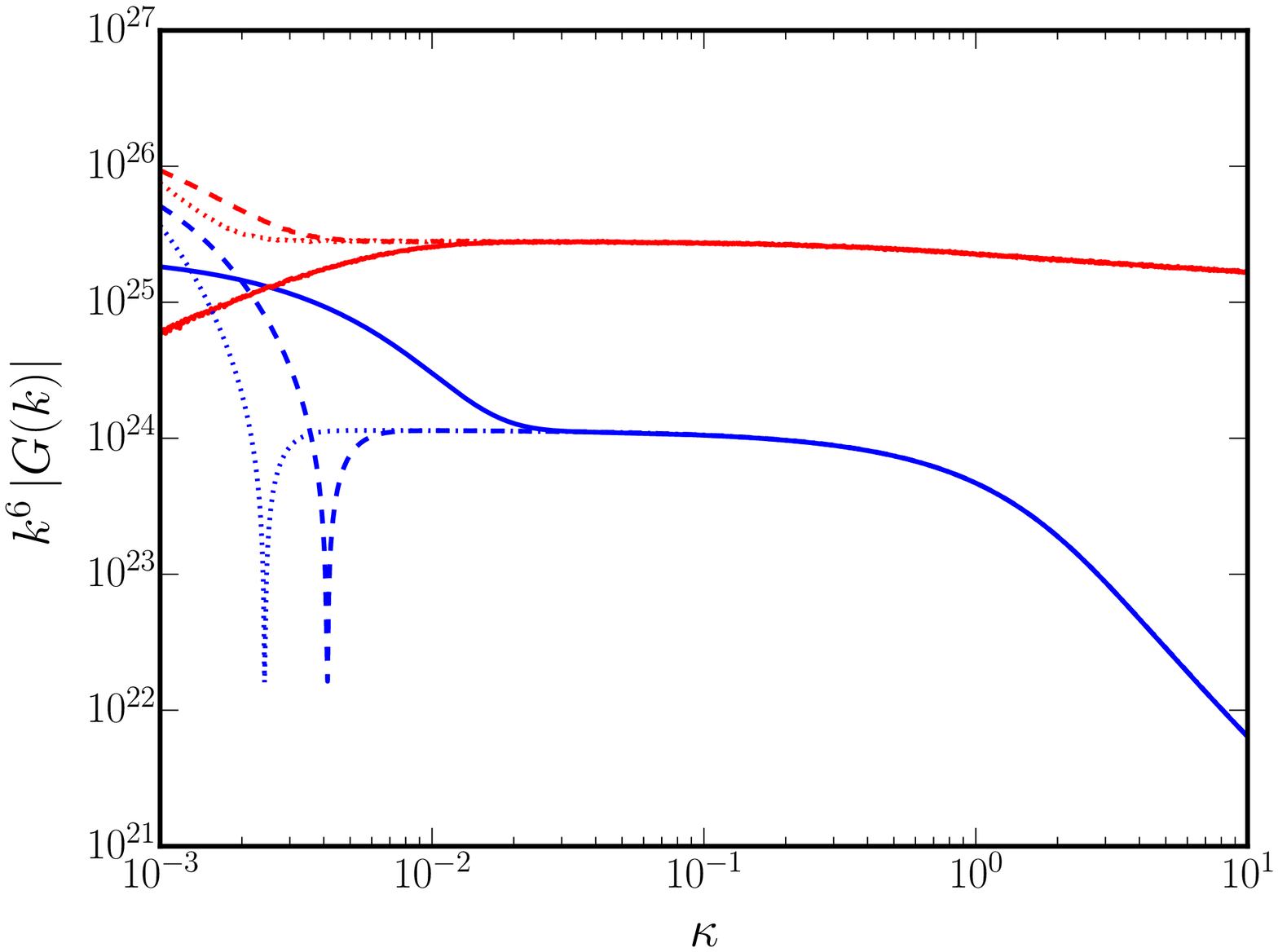}\\
\includegraphics[height=6.00cm,width=7.50cm]{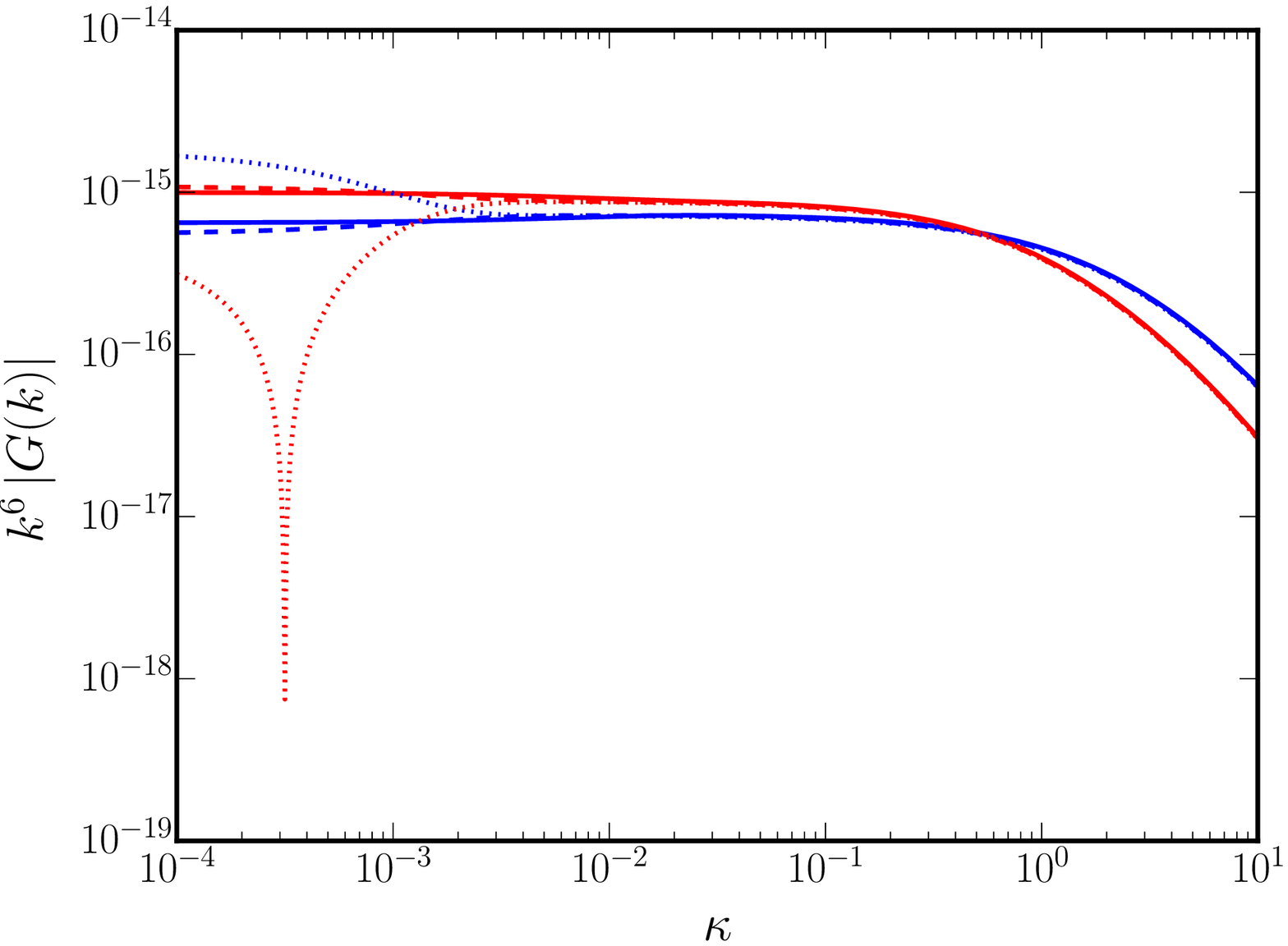}
\includegraphics[height=6.00cm,width=7.50cm]{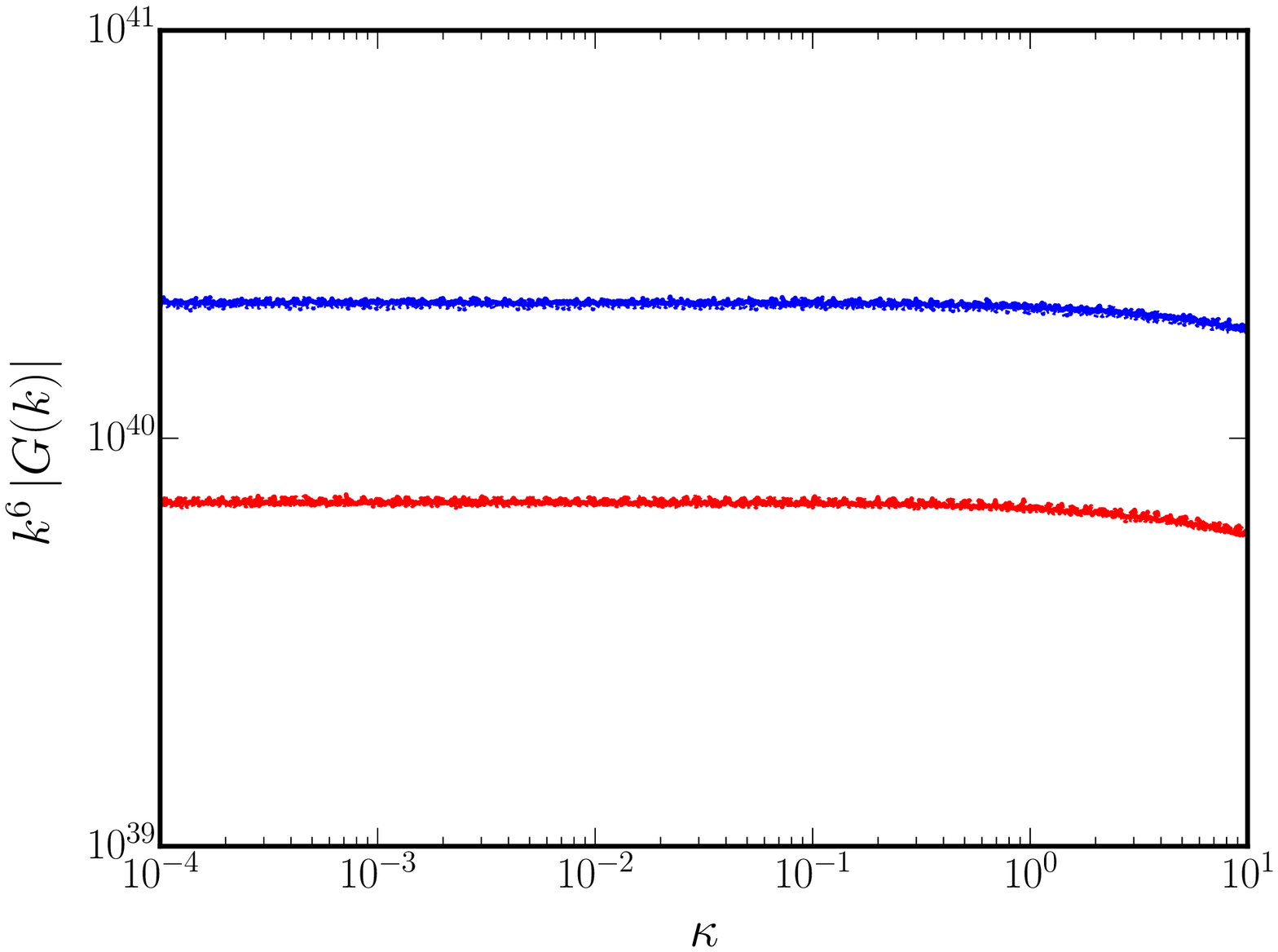}
\end{center}
\caption{The amplitudes of the contributions to the three-point function 
arising from the first (in blue) and second (in red) integrals 
[cf.~Eqs.~(\ref{eq:int-h-eta})] have been plotted for the $n=1$ case (on 
the left) and the $n=2$ case (on the right) as functions of the cut-off 
parameter $\kappa$.
We have plotted the results for three different choices of $N_i$ corresponding 
to $k=100\,\sqrt{J''/J}$ (as solid lines), $k=500\,\sqrt{J''/J}$ (as dashed
lines) and $k=1000\,\sqrt{J''/J}$ (as dotted lines).
Also, we have plotted them for the non-helical case (on top) as well 
as the helical case with $\gamma=2$ (at the bottom).
Note that the two contributions to the three-point function have been 
multiplied with suitable powers of $k$.
It should be clear from the above plots that, for the choice of $k=300\,
\sqrt{J''/J}$, the resulting contributions are largely insensitive to 
$\kappa$ around $\kappa=0.1$.
Because of this reason, we shall choose to evaluate the three-point 
function and the corresponding non-Gaussianity parameter (see the 
next section) for this choice of values.
It is also useful to note from the final plot that the three-point function 
corresponding to $n=2$ in the helical case is largely unaffected by the 
value of the sub-Hubble cut-off parameter.
As we shall see in the next figure, this occurs due to the fact that the 
dominant contribution to this three-point function arises from the 
super-Hubble domain.}\label{fig:kappa}
\end{figure}

\begin{figure}[!t]
\begin{center}
\includegraphics[height=6.00cm,width=7.50cm]{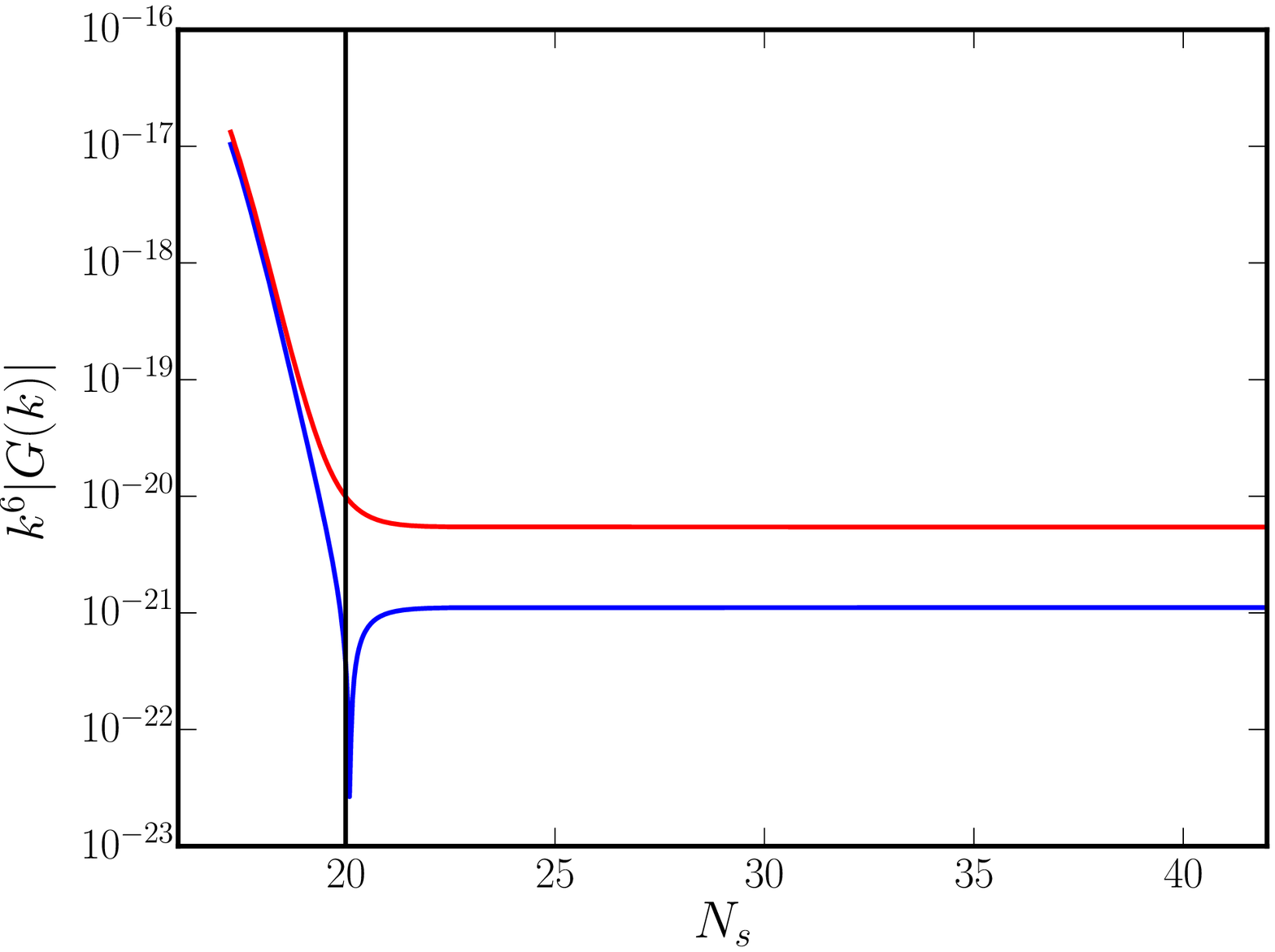}
\includegraphics[height=6.00cm,width=7.50cm]{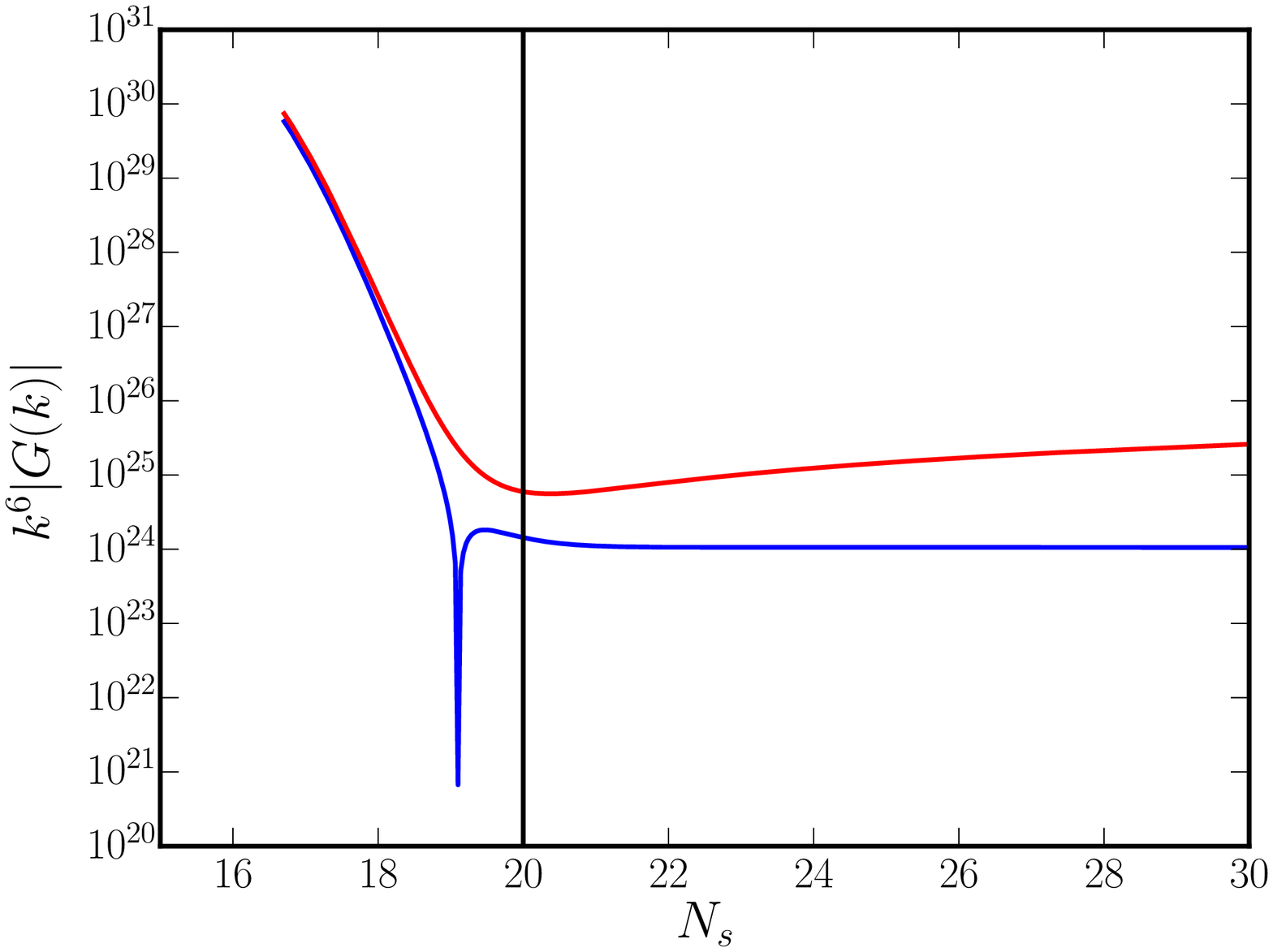}\\
\includegraphics[height=6.00cm,width=7.50cm]{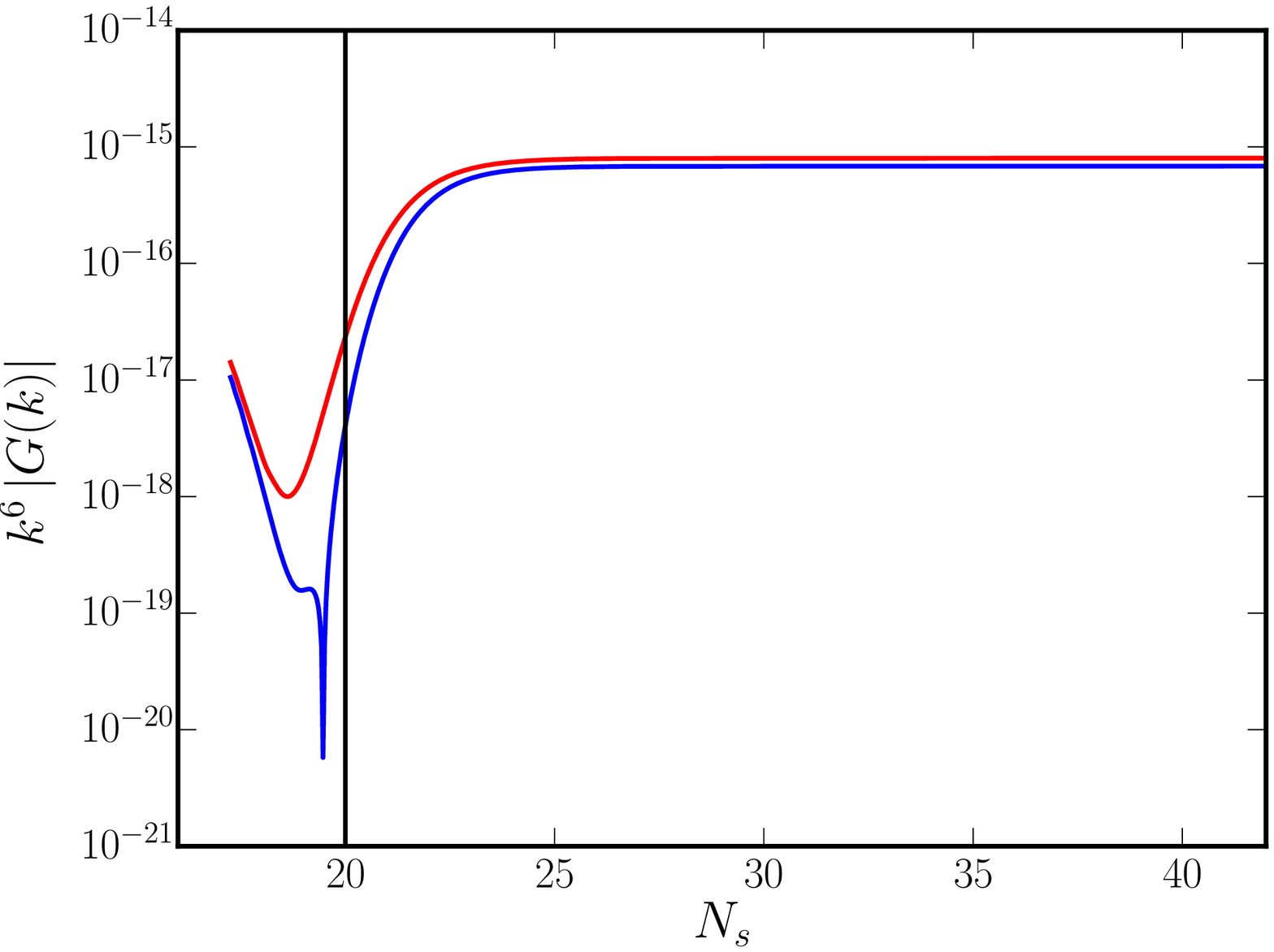}
\includegraphics[height=6.00cm,width=7.50cm]{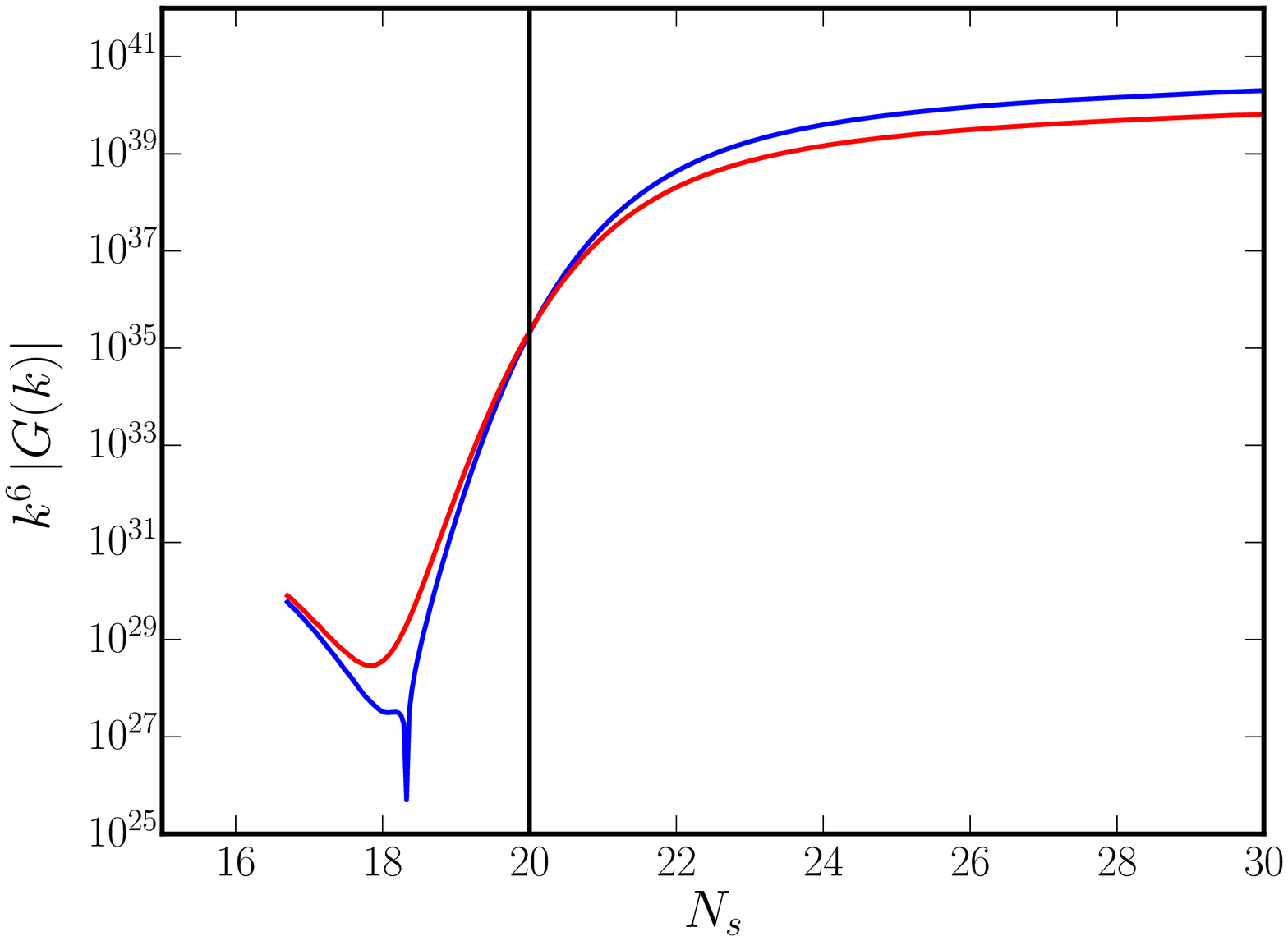}
\end{center}
\caption{The amplitudes of the contributions to the three-point function 
arising from the first and the second integrals have been plotted for the 
$n=1$ and the $n=2$ cases as functions of $N_s$, with the same choice of 
colors as in the previous figure.
Also, we have plotted the non-helical and the helical cases just as in the 
last figure.
We have imposed the initial condition at an $N_i$ corresponding to $k=300\,
\sqrt{J''/J}$, and we have set $\kappa=0.1$ in arriving at the above plots.
We have plotted the results for a mode with the wavenumber $k=0.002\,
{\rm Mpc}^{-1}$, which is often chosen as the pivot scale when comparing
with the observations.
The black vertical lines in the plots indicate the e-fold at which the 
mode leaves the Hubble radius.
Note that, in the $n=1$ case, the amplitude of the two terms freeze soon 
after Hubble exit, which implies that the super-Hubble contributions to 
the three-point function are negligible.
This is true in the corresponding helical case as well.
When $n=2$, for the non-helical case, even though the first integral 
flattens out in the super-Hubble limit, the contribution due to the 
second integral continues to grow. 
This behavior has been encountered in analytical calculations and it can 
be attributed to the ${\rm ln}(-k\,\ee)$ term that arises in this case.
It is also clear from the plots that the introduction of helicity 
considerably enhances the amplitude of the three-point functions.
Moreover, the enhancement occurs around the time the mode leaves the 
Hubble radius.
This is further accentuated by the super-Hubble contributions that are
encountered in the $n=2$ case.}\label{fig:nshs}
\end{figure}

In order to identify suitable values of $\kappa$, $N_i$ and $N_s$, we 
shall evaluate the two contributions to the three-point function of our 
interest [described by the  integrals~(\ref{eq:int-h-eta})] in the 
equilateral limit for different combinations of these variables. 
Earlier, we had mentioned that we solve the differential equations
involved [\viz Eqs.~(\ref{eq:Asigma-diff-N}) and~(\ref{eq:fk-diff-N})]
using the fifth order Runge-Kutta routine.
We shall carry out the integrals~(\ref{eq:int-h-eta}) using the Boole's 
rule~\cite{Press:2007:NRE:1403886}.
We should add that, in our discussion hereafter, 
for convenience, we shall simply set the polarization tensor 
$\varepsilon^{\bm{k}}_{\sigma i}$ to unity.
However, we shall include all the contributions due to the positive
and negative helicity modes as well as the cross terms that arise.
We shall first keep $N_s$ fixed and calculate the quantities for three 
different values of $N_i$ and varied $\kappa$.
This helps us identify a suitable combination of $N_i$ and $\kappa$ for 
which the three-point function is insensitive to the choice of these 
variables.
We shall then choose these values of $N_i$ and $\kappa$ and attempt to 
identify a suitable choice for $N_s$. 
The results of these exercises are plotted in Figs.~\ref{fig:kappa} 
and~\ref{fig:nshs} for both the non-helical and helical cases for 
$n=1$ as well as $n=2$.
In Fig.~\ref{fig:kappa}, we have illustrated the dependence of the two
contributions to the three-point function on the sub-Hubble cut-off 
parameter $\kappa$ for different choices of $N_i$.
It is clear from the figure that, for instance, for $N_i$ corresponding
to $k=300\,\sqrt{J''/J}$, the two contributions to the three-point
function are largely independent of $\kappa$ around $\kappa=0.1$.
Therefore, we shall work with these values.

\par

Let us now turn to determining a suitable $N_s$.
With $\kappa$ and $N_i$ fixed at the aforementioned values, in 
Fig.~\ref{fig:nshs}, we have plotted the two contributions as a function 
of $N_s$.
Note that, in the $n=1$ case, the results are independent of the choice
$N_s$, provided we choose it corresponding to a time reasonably after 
Hubble exit.
In contrast, when $n=2$, it is clear from the figure that there is a slow growth
as a function of $N_s$.
Such a behavior is peculiar to the model and the choices of the parameter 
involved.
This growth is well known from the analytical calculations and, in the 
non-helical case, it can be shown to behave as ${\rm ln}\,(-\,k\,\ee)$
(in the equilateral limit we are focusing on), which is exactly the 
behavior we observe numerically~\cite{Jain:2012vm,Chowdhury:2018blx}.
It is also clear from Fig.~\ref{fig:nshs} that the introduction of helicity 
considerably enhances the amplitude of the three-point function in both 
the $n=1$ and $n=2$ cases.
Moreover, it is evident that the enhancement occurs as the modes leave 
the Hubble radius, which is further accentuated in the $n=2$ case due 
to the super-Hubble contributions that arise.
While the results will be independent of $N_s$ in the $n=1$ case, they
will strongly depend on the parameter when $n=2$.
Ideally, it would have been desirable to evaluate the three-point function 
when inflation ends at $60$ e-folds or so since the earliest time when, say, 
the largest scale had exited the Hubble radius. 
However, evolving the modes for this duration and evaluating the integrals
introduce inaccuracies after about $30$ e-folds or so.
Therefore, we shall evaluate the three-point function in the $n=2$ case for
$N_s \sim 30$.

\section{Amplitude and shape of the non-Gaussianity 
parameter}\label{sec:bnl}

In this section, we shall introduce the dimensional non-Gaussianity
parameter $\bnl$ which characterizes the amplitude and shape of the 
three-point function involving the helical magnetic field and the 
perturbations in the scalar field.
We shall present the numerical results for $\bnl$ for the different
cases we had discussed.
In order to illustrate the accuracy of our numerical methods, we shall 
also compare our numerical results with the analytical results that
are available in the non-helical case.

\par

The amplitude of the non-Gaussianity in the local model 
for the scalar three-point function is usually parameterized in terms 
of a parameter $\fnl$ which, in this model, coincides with the bispectrum 
scaled by products of the power spectra.
Here, we generalize that analogously to a non-Gaussianity parameter 
$\bnl$ which we define through the following relation:
\begin{eqnarray}
\hat{B}_{i\,\bm{q}}(\eta) 
= \hat{B}_{i\,\bm{q}}^{({\rm G})}(\eta) 
+ \frac{\bnl}{2\,\Mp}\, 
\int \f{\d^3\bm{p}}{(2\,\pi)^{3/2}}\,
\hat{\delta\phi}_{\bm{q}-\bm{p}}(\eta)\,
\hat{B}_{i\,{\bm p}}^{({\rm G})}(\eta),
\end{eqnarray}
where $\hat{B}_{i\,\bm{q}}$ is the Fourier mode of the actual magnetic 
field and $\hat{B}_{i\,\bm{q}}^{({\rm G})}$ indicates the Fourier mode of
its Gaussian part, while, as usual, $\hat{\delta\phi}_{\bm{q}-\bm{p}}$ 
refers to Fourier mode of the perturbation in the scalar field which 
has already been assumed to be Gaussian.
Upon using Wick’s theorem to calculate the three-point function
of our interest, we find that we can express the parameter $\bnl$
as
\begin{eqnarray}\label{eq:bnl}
\bnl\l(\bm{k_1},\bm{k_2},\bm{k_3}\r) 
&=& \f{1}{16\,\pi^5}\, \f{J^2(\ee)}{a^2({\ee})}\,
\biggl[\ka^3\,\kb^3\,\kc^3\,
G_{\delta\phi B B}(\bm{k_1},\bm{k_2},\bm{k_3})\biggr]\nn\\
& &\times\, \biggl\{\cP_{\delta\phi}(\ka)\left[\kc^3\,\psb(\kb)
+ \kb^3\,\psb(\kc)\right]\biggr\}^{-1}.
\end{eqnarray}
In this expression, $\psb(k)$ denotes the power spectrum of the magnetic 
field that we have discussed earlier, while $\cP_{\delta\phi}(k)$ denotes 
the power spectrum of the perturbations in the scalar field that is defined 
as
\begin{equation}
\cP_{\delta\phi}(k) 
= \f{k^3}{2\,\pi^2\,\Mp^2}\,\vert\,f_{k}(\ee)\vert^2,
\end{equation}
with the right hand side to be evaluated as $\ee\to 0$.

\par

\begin{figure}[!t]
\begin{center}
\includegraphics[width=7.50cm]{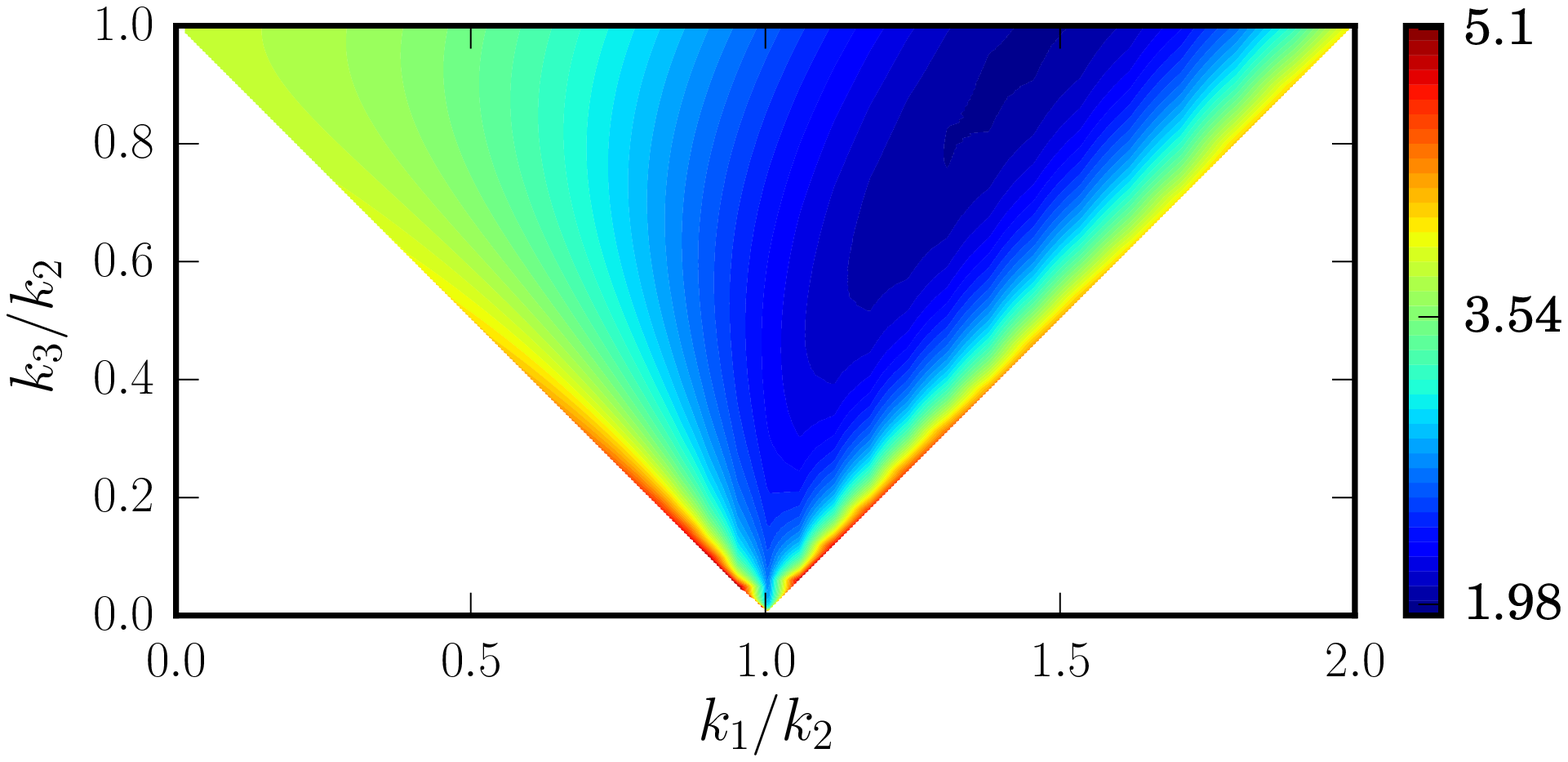}
\includegraphics[width=7.50cm]{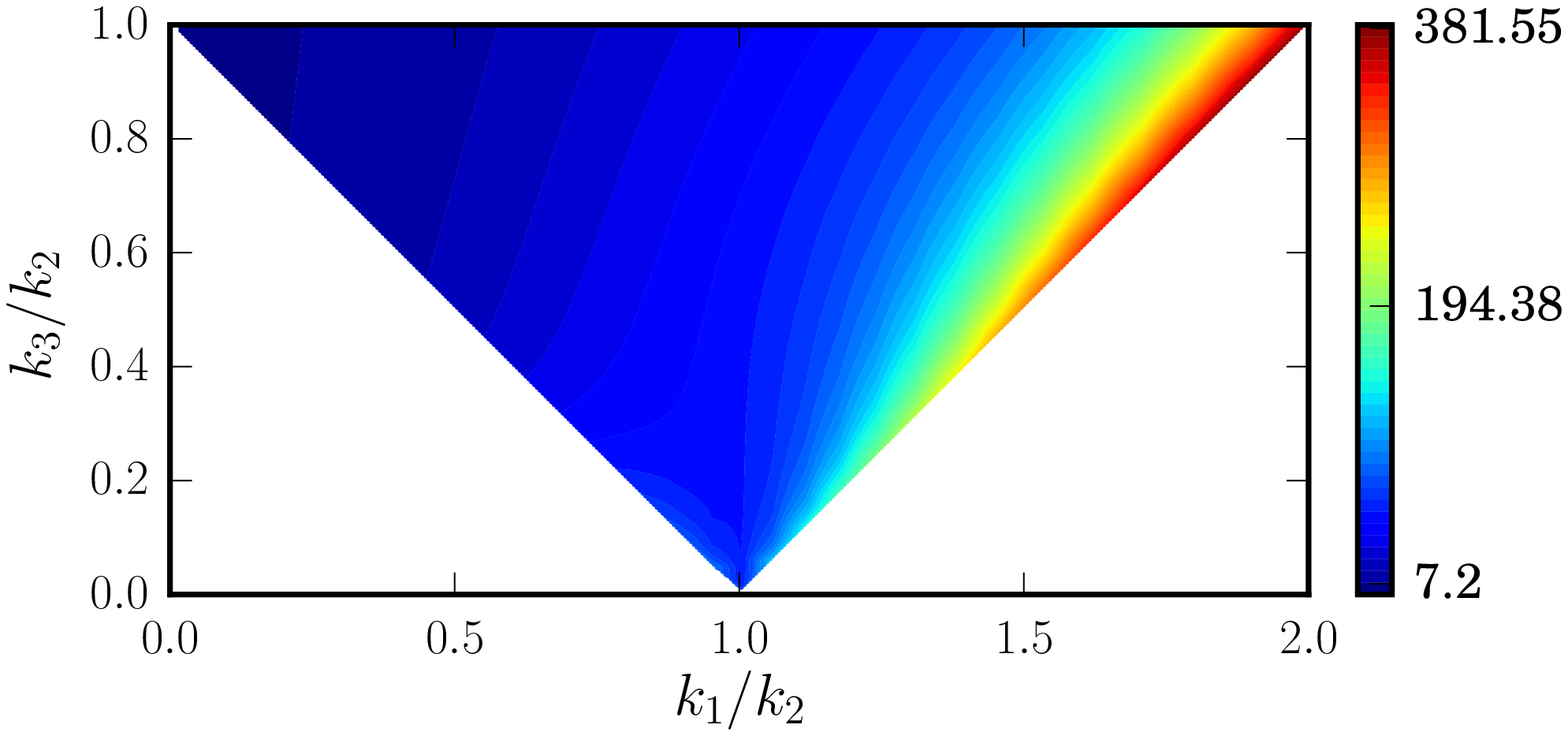}\\
\includegraphics[width=7.50cm]{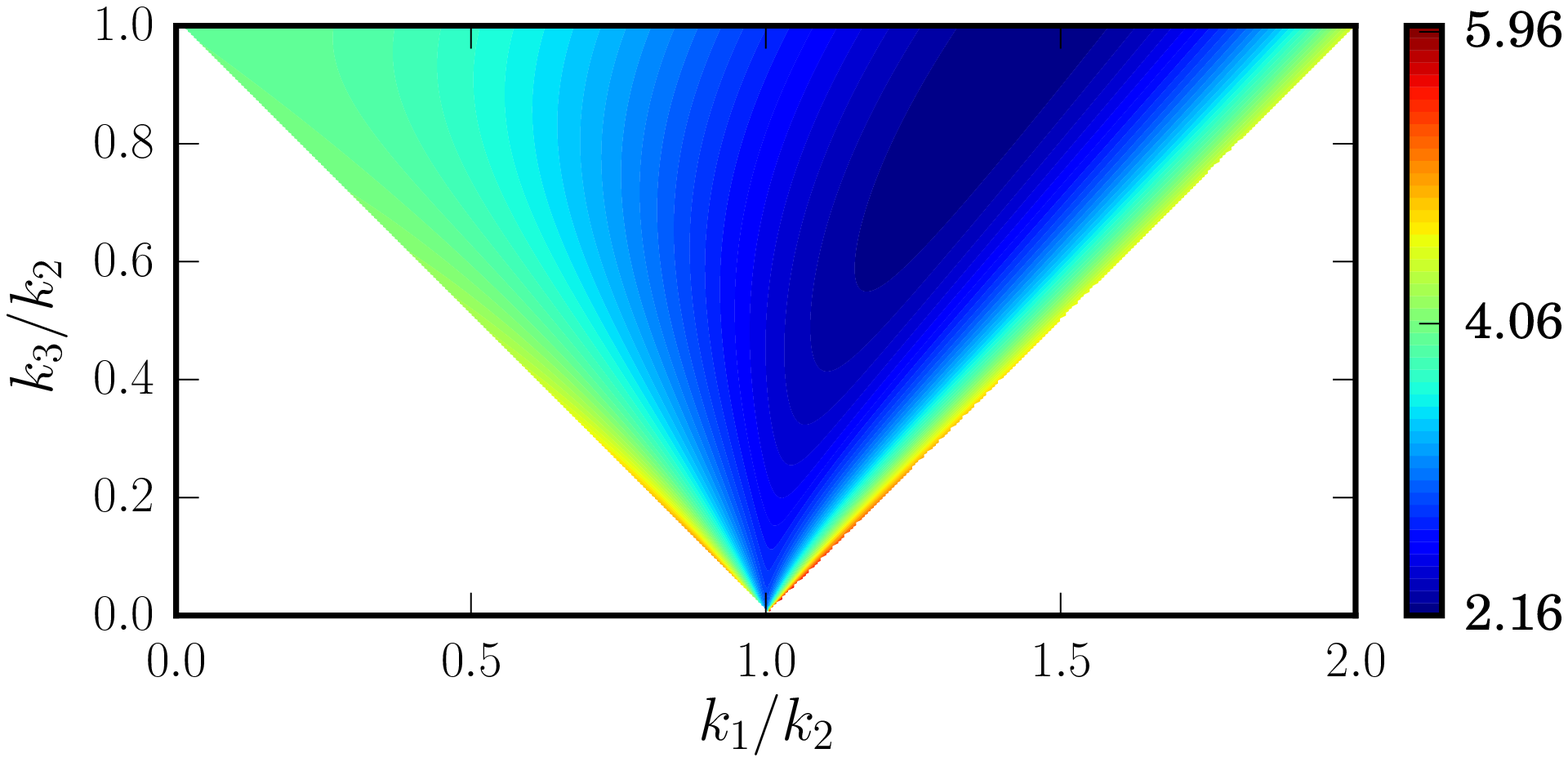}
\includegraphics[width=7.50cm]{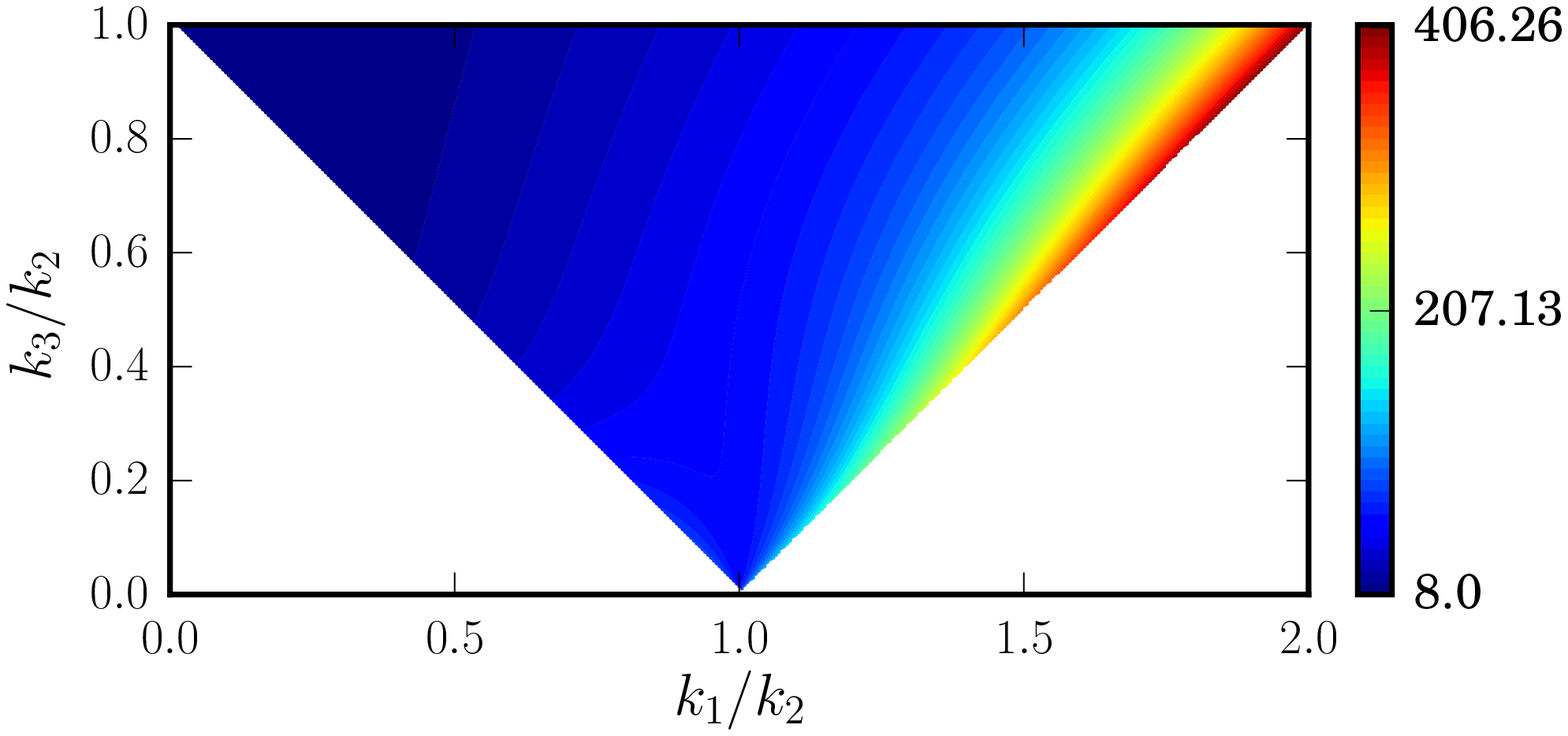}\\
\includegraphics[width=7.50cm]{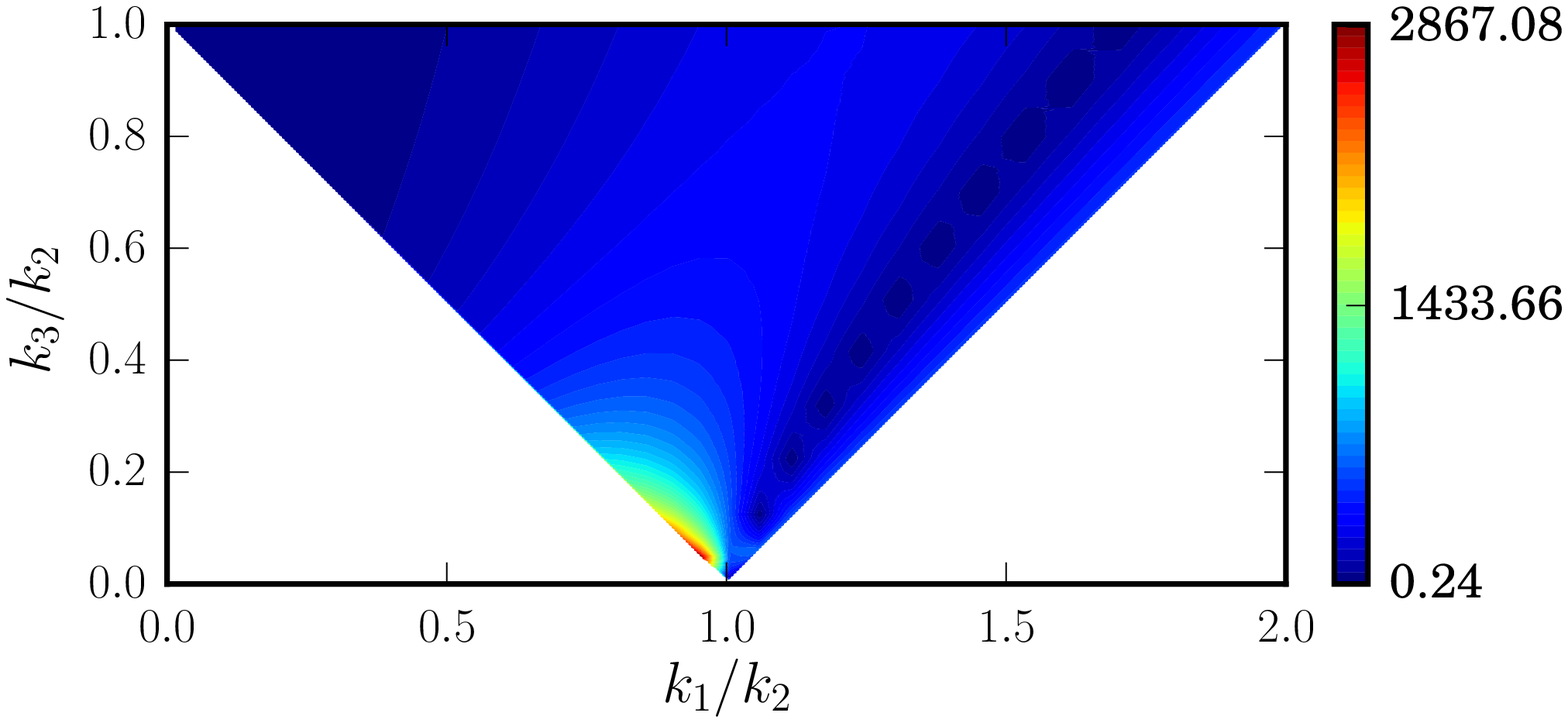}
\includegraphics[width=7.50cm]{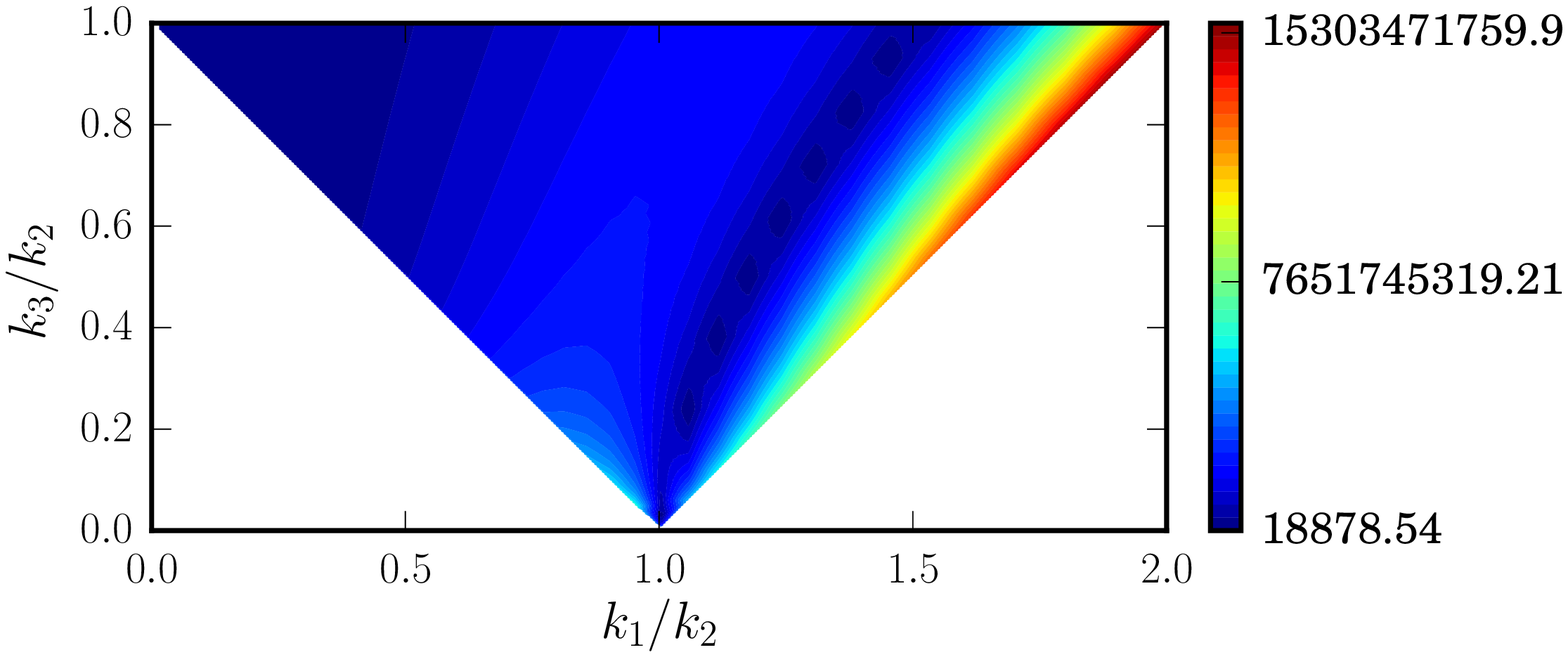}
\end{center}
\caption{The non-Gaussianity parameter $\bnl$ has been plotted for the 
non-helical case obtained numerically (top row), analytically (middle 
row), and the helical case arrived at numerically (third row of plots) 
for the cases of $n=1$ (on the left) and $n=2$ (on the right).
For reasons we have discussed, in the $n=2$ case, we have evaluated the
non-Gaussianity parameter corresponding to $N_s \sim 30$.
From the first and second rows, we find that the analytical and numerical 
results match up to $5$--$10\%$.
It is evident from the third row of plots that the helical term substantially 
boosts the value of $\bnl$ for both the $n=1$ and $n=2$ cases, with the 
amplification being larger for $n=2$.}\label{fig:bnl}
\end{figure}

In Fig.~\ref{fig:bnl}, we have plotted the numerical results for the 
non-Gaussianity parameter $\bnl$ (for an arbitrary triangular 
configuration of wavenumbers, in this context, see 
Ref.~\cite{Chowdhury:2018blx}) for the cases of $n=1$ and $n=2$ when 
$\gamma=0$ and $\gamma=2$.
We have also plotted the analytical results available for $n=1$ and 
$n=2$ in the non-helical case, to illustrate the extent of accuracy
of our numerical methods.
On comparing the results in the non-helical case for $n=1$ and $n=2$, 
we find that the analytical and numerical results match up to about 
$5$--$10\%$.
Recall that we have been working with $\gamma=2$.
Even for such a relatively small value of $\gamma$, we find that the
introduction of helicity considerably amplifies non-Gaussianities.


\section{Summary}\label{sec:c}

The generation and evolution of primordial magnetic fields has been 
studied extensively in the context of inflation.
By introducing a non-minimal coupling term in the standard electromagnetic 
action, it has been possible to obtain scale invariant magnetic fields of 
the requisite amplitude to be in conformity with observations.
It has also been realized that adding a parity violating term to the 
electromagnetic action results in the production of helical magnetic 
fields, which can have significant observational 
imprints~\cite{Caprini:2003vc,Ballardini:2014jta,Seto:2008sr,Durrer:2010mq}.
Specifically, helicity considerably boosts the amplitude of the power 
spectrum of the magnetic field.

\par

The non-Gaussian signatures generated via cross-correlations between the 
primordial magnetic fields and the perturbations in a scalar field can 
provide additional constraints to characterize the magnetic fields.
These non-Gaussianities, at the level of three-point functions, have been 
examined earlier in the non-helical case~\cite{Caldwell:2011ra,Motta:2012rn,
Jain:2012ga,Jain:2012vm}.
In this work, we numerically evaluate the three-point functions involving 
primordial helical magnetic fields and the perturbations in a scalar field.
Since the introduction of the helical term in the action amplifies the 
strength of the magnetic field, it can also be expected to lead to larger 
non-Gaussianities.
We find that even with a small value of the parameter quantifying the extent 
of helicity, there is a substantial enhancement in the non-Gaussianity 
parameter~$\bnl$. 
It would be interesting to explore related observational consequences, and 
possibly arrive at constraints on the mechanisms that could lead to the 
generation of the helical magnetic fields (in this context, see Ref.~\cite{Kunze:2013hy}).
Another important aspect would be to investigate into the behavior of 
the three-point functions involving the helical magnetic fields and 
the curvature perturbation~\cite{Motta:2012rn,Jain:2012vm}.
Importantly, in such a case, one may encounter additional terms arising
in the action describing the interaction term.
Also, while the magnitude of $\bnl$ in, say, slow roll inflation, may not 
differ substantially from the case we have examined, considering the curvature 
perturbation can provide us with additional parameters (\viz the slow roll 
parameters) to more effectively constrain the non-Gaussianities generated.
We are presently working on these issues.


\section*{Acknowledgements}

This work was initiated during visits by DC and LS to the Department 
of Physics and Astronomy, Johns Hopkins University, Baltimore, Maryland, 
U.S.A. under the aegis of The Indo-US Science and Technology Forum 
grant~IUSSTF/JC-Fundamental Tests of Cosmology/2-2014/2015-16.
DC would like to thank the Indian Institute of Technology Madras, Chennai, 
India, for financial support through half-time research assistantship.
LS also wishes to thank the Indian Institute of Technology Madras, 
Chennai, India, for support through the Exploratory Research Project
PHY/17-18/874/RFER/LSRI.
This work was supported at Johns Hopkins University by NSF
Grant No. 1519353, NASA NNX17AK38G, and the Simons Foundation.
We wish to thank Rajeev Jain for comments on the manuscript.

\bibliographystyle{JHEP}
\bibliography{hmf-3pt-infl-october-2018}
\end{document}